\documentclass[twocolumn]{aastex63}

\usepackage{natbib}
\usepackage{CJK}

\usepackage{natbib}
\usepackage{graphicx}
\usepackage{amsmath}
\usepackage{booktabs}
\usepackage{longtable}
\usepackage{xspace}
\usepackage{hyperref}
\usepackage[T1]{fontenc}
\usepackage{lipsum}
\usepackage{comment}
\usepackage{xcolor}
\usepackage{chngcntr}
\usepackage{textcomp}
\newcommand{\ttilde}{\raisebox{0.5ex}{\texttildelow}}

\received{xxx}
\revised{xxx}
\accepted{xxx}

\shorttitle{Sample article}
\shortauthors{Zhang et al.}

\graphicspath{{./}{figures/}}
\begin{document}
\begin{CJK*}{UTF8}{gbsn}
\title{Discovery of a Jupiter Analog Misaligned to the Inner Planetary System in HD 73344}

\correspondingauthor{Jingwen Zhang}
\email{jingwen7@hawaii.edu}

%\author[0000-0002-0786-7307]{Greg J. Schwarz}
%\affiliation{Institute for Astronomy, University of Hawaii at Manoa\\2680 Woodlawn Dr. \\Honolulu, HI 96822, USA}

\author[0000-0002-2696-2406]{Jingwen Zhang (张婧雯)}
\altaffiliation{NASA FINESST Fellow}
\affiliation{Institute for Astronomy, University of Hawaiʻi at M\=anoa, 2680 Woodlawn Drive, Honolulu, HI 96822, USA}

\author[0000-0002-3725-3058]{Lauren M. Weiss}
\affiliation{Department of Physics and Astronomy, University of Notre Dame, Notre Dame, IN 46556, USA}

\author[0000-0001-8832-4488]{Daniel Huber}
\affiliation{Institute for Astronomy, University of Hawaiʻi at M\=anoa, 2680 Woodlawn Drive, Honolulu, HI 96822, USA}
\affiliation{Sydney Institute for Astronomy (SIfA), School of Physics, University of Sydney, NSW 2006, Australia}

\author[0000-0002-6618-1137]{Jerry W. Xuan}
\altaffiliation{NASA FINESST Fellow}
\affiliation{Department of Astronomy, California Institute of Technology, Pasadena, CA 91125, USA}

\author[0000-0003-1341-5531]{Michael Bottom}
\affiliation{Institute for Astronomy, University of Hawaiʻi at M\=anoa, 640 N. Aohoku Pl, Hilo, HI 96720, USA}

\author[0000-0003-3504-5316]{Benjamin J.\ Fulton}
\affiliation{NASA Exoplanet Science Institute/Caltech-IPAC, California Institute of Technology, Pasadena, CA
91125, USA}

\author[0000-0002-0531-1073]{Howard Isaacson}
\affiliation{Department of Astronomy, 501 Campbell Hall, University of California, Berkeley, CA 94720, USA}

\author[0000-0003-2562-9043]{Mason G.\ MacDougall}
\affiliation{Department of Physics \& Astronomy, University of California Los Angeles, Los Angeles, CA 90095, USA}

\author[0000-0003-2657-3889]{Nicholas Saunders}
\altaffiliation{NSF Graduate Research Fellow}
\affiliation{Institute for Astronomy, University of Hawaiʻi at M\=anoa, 2680 Woodlawn Drive, Honolulu, HI 96822, USA}

%\collaboration{1}{(AAS Journals Data Scientists collaboration)}
%\collaboration{1}{(LaTeX collaboration)}
%\nocollaboration{2}

\begin{abstract}

We present the discovery of a Jupiter-like planet, HD 73344 d ($m_{d}=2.55^{+0.56}_{-0.46}\ \mathrm{M_{J}}$, $a_{d}=6.70^{+0.25}_{-0.26}$ AU, $e_{d}=0.18^{+0.14}_{-0.12}$) based on 27-year radial velocity observations from ELODIE, Lick/Hamilton, SOPHIE, APF and Keck/HIRES. HD 73344 also hosts a compact inner planetary system, including a transiting sub-Neptune HD 73344 b ($P_{b}=15.61\ \mathrm{days}$, $r_{b}=2.88^{+0.08}_{-0.07}\ \mathrm{R_{\oplus}}$) and a non-transiting Saturn-mass planet ($P_{c}=65.936\ \mathrm{days}$, $m_{c}\sin{i_c}=0.367^{+0.022}_{-0.021}\ \mathrm{M_{J}}$).  By analyzing  \textit{TESS} light curves, we identified a stellar rotation period of $9.03\pm{1.3}$ days. Combining this with $v\sin{i_*}$ measurements from stellar spectra,  we derived a stellar inclination of $63^{\circ}.6^{+17.4}_{-16.5} $. Furthermore, by combining radial velocities and Hipparcos-Gaia astrometric acceleration, we characterized the three-dimensional orbit of the outer giant planet and constrained its mutual inclination relative to the innermost transiting planet to be $46^{\circ} <\Delta I_{bd}< 134^{\circ}\ (1\sigma)$ and $20^{\circ} <\Delta I_{bd}< 160^{\circ}\ (2\sigma)$, strongly disfavoring coplanar architectures. Our analytical calculations and N-body simulation reveal that the two inner planets are strongly coupled with each other and undergo nodal precession together around the orbital axis of the giant planet. During nodal precession, the orbital inclination of inner planets oscillate with time and therefore become misaligned relative to the stellar spin axis. The formation of such systems suggests a history of planet-planet scattering or misalignment between the inner and outer components of protoplanetary disks. The upcoming release of Gaia DR4  will uncover more systems similar to HD 73344 and enable the study of the flatness of exoplanet systems with a mixture of inner and outer planetary systems on a statistical level.

\end{abstract}

%, which suggests that the transiting planet HD 73344 b may be misaligned with the stellar spin axis in the line-of-sight dimension with $1\sigma$ significance.
%% Keywords should appear after the \end{abstract} command. 
%% See the online documentation for the full list of available subject
%% keywords and the rules for their use.
\keywords{editorials, notices --- 
miscellaneous --- catalogs --- surveys}

\section{Introduction}\label{sec:intro}

One key question in exoplanet science is how outer giant planets dynamically influence the formation and evolution of inner small planets (e.g. super-Earths and sup-Neptunes) in multi-planetary systems. Jupiter is believed to have played a decisive role in the dynamical evolution of the early Solar System, by shaping distribution of raw materials for the formation of terrestrial planets \citep{walsh2011} and shepherding water-bearing comets to inner planetary systems \citep{OBrien2018}. Numerous theoretical studies have explored the potential influence of outer giant planets on  inner planet formation. Some propose that the migration of giant planets could aid in transporting materials to inner orbits and enhancing the formation of inner planetary systems \citep{Mandell2007}. \cite{Best2024} proposed that interactions between giant planets and disks may gradually transport rings of planetesimals into inner planetary systems to support the formation of inner planets.

In exoplanet systems, the connection between small, inner planets and outer giant planets is a topic of ongoing debate.  Some studies suggest that the presence of cold giant planets ($>1\mathrm{AU}$, $Mp > 0.5 M_{J}$) may enhance the probability of finding inner small planets (1-4 $R_{\oplus}$, 1-10 $ M_{\oplus}$) in the same system (e.g. \citealt{ZW2018, Bryan2019, Rosenthal2022}). 
However, recent studies based on new and statistically homogeneous samples point to a lack of correlation between outer giant planets and inner transiting planets.  \cite{Bonomo2023} examined 37 Kepler and K2 systems with close-in small planets using RV observations from HARPS-N and obtained an occurrence rate of $9.3^{+7.7}_{-2.9}\%$ for cold Jupiters between $1-10$ AU around systems with inner small planets, suggesting a neutral or even negative correlation.  The Kepler Giant Planet Survey of 63 Kepler systems with small transiting planets, which netted 13 Jupiter-mass non-transiting companionsn has also added to the discussion \citep{Weiss2024}.  Incorporating the new data, \cite{Zhu2024} and \cite{Bryan2024} find no strong correlation between inner super-Earths and outer giants in the population as a whole, but do find an enhancement of cold Jupiter companions to the super-Earths that orbit metal-rich stars ([Fe/H] $> 0$, $2\sigma$ significance).     %To address the discrepancy, \cite{Zhu2024} and \cite{Bryan2024} argues that interpreting this correlation requires consideration of the host star metallicity distribution. They found that the strong correlation between inner small planets and outer giant planets holds true primarily in metal-rich stars. %\cite{ZW2018} and \cite{Bryan2019} independently  analyzed the samples of  known close-in small planets  using public radial velocity (RV) observations. Both studies reported a excess rate of cold Jupiters ( $32\%\pm{8}\%$ and $39\%\pm{7}\%$) around stars hosting close-in small planets compared to occurrence of cold Jupiters around field stars ($20.2^{+6.3}_{-3.4}\%$, \citealt{Wittenmyer2020}).
%\cite{Rubenzahl_2021} studied a homogeneous sample from the California Plane Search (CPS) survey using Keck/HIRES and found that $41^{+13}_{-15}\%$ of systems containing  close-in small planets also host at least an outer giant planet. In contrast, the cold Jupiters occurrence rate irrespective of small planets presence from the same survey was found to be $17.6^{+1.9}_{-2.4}\%$. %based on new and statistically homogeneous samples %The Kepler Giant Planet Survey of 63 Kepler systems with small transiting planets, which netted 13 Jupiter-mass non-transiting companions, and for which detailed completeness calculations are still in progress, will add to the discussion (\citealt{Weiss2024}, Weiss et al. in prep.). 

%However, it is unclear whether the positive correlation arises from the fact that massive disks have larger reservoir of solids to nucleate both small planets and outer giants, or if the presence of outer giant planets facilitates the formation of inner small planets. 

Another important aspect of how outer giant planets influence inner small planets is the coplanarity of the multi-planetary systems. In the Solar System, planets orbit within only a few degrees of the solar equatorial plane, believed to result from the protoplanetary disk where these planets formed. On the other hand, it has been discovered that giant planets could exhibit significant misalignment relative to the inner planets, e.g. $\pi$ Men and HAT-P-11 systems \citep{Xuan2020, DeRosa2020,Damasso2020}. The misalignment could be the result of dynamical events. One possible scenario is that all planets are initially aligned with each other in the protoplanetary disk. Later, interactions such as planet-planet scattering \citep{Rasio1996,Chatterjee2008,Beaug2012, Petrovich2014} or disturbances from fly-by stars \citep{Malmberg2011} could misalign the orbits of giant planets out of the original disk plane. Subsequently, when the outer giant planet becomes misaligned relative to inner super-Earths or sub-Neptunes, it could tilt the orbits of the inner planets and cause them to be misaligned relative to their host stars \citep{Huber2013, DongandPu2017,zhang2021}. Alternatively, the planets could formed in a warped protoplanetary disk and thus misaligned when they formed. An intriguing pattern that has emerged is that among the systems with multiple transiting planets, the gap complexity of the transiting planets is a strong predictor of the presence of an outer giant planet, suggesting that the giant planet might dynamically heat the inner system in a manner that leads to collisions and/or ejections \citep{He2023}.  Measurements of the mutual inclinations of all the bodies in multi-planet system can distinguish between these  dynamical histories.

One way to measure the mutual inclination between inner and outer planets is to characterize the three-dimensional orbits of giant planets outside inner transiting planets by combining radial velocities (RVs) and Hipparcos-Gaia astrometric astrometric acceleration \citep{Kervella2019,Brandt2021}. However, the measurements are still limited to a small number of systems because the methods requires both long-term RV observations and a significant astrometric acceleration. Here, we present the system HD 73344, which is a transiting planet host with significant astrometric acceleration. The system hosts a transiting sub-Neptune, HD 73344 b ($P_b=15.61$ days, $r_{b}=2.88^{+0.08}_{-0.07}\ \mathrm{R}_{\earth}$), discovered by the \textit{K2} mission \citep{Yu2018}. \textit{TESS} and \textit{Spitzer} observations subsequently confirmed the transiting signal. \cite{Sulis2024} measured an upper limit on the mass of the transiting planet as $m_{b}<10.48\ \mathrm{M}_{\earth}\ (3\sigma)$ using high-precision RVs from SOPHIE and HIRES. With the same RV dataset, they also reported the discovery of a non-transiting Saturn-mass planet, HD 73344 c ($m_{c}\sin{i_c}=116.3^{+12.8}_{-13.0}\ \mathrm{m}_{\earth}$), at an orbital period of $\ttilde 66$ days. In this paper, we report the discovery of a long-period giant planet ($m_{d}=2.55^{+0.56}_{-0.46}\ \mathrm{M_{J}}$, $a_{d}=6.70^{+0.25}_{-0.26}$ AU) outside the two  inner planets based on 27 years RV observations. Moreover, we characterize the three-dimensional orbital parameters of this outer planet by combining RVs and Hipparcos and Gaia astrometric data, and constrain the mutual inclination between the outer giant planet and the innermost transiting planet.

\section{Observations} \label{sec:obs}

\subsection{ELODIE Radial Velocities}

HD 73344 was observed by the ELODIE spectrograph \citep{Baranne1996}, installed on the 1.93m  telescope located at the Observatoire de Haute Provence, France. ELODIE was a fibre-fed echelle spectrograph with a resolution power of 42 000 and a wavelength range from 389.5 to 681.5 nm split into 67 spectral orders. We analysed the 73 spectra acquired between November 1997 and March 2006 which have a median signal-to-noise ratio of 85 at 555 nm, and are publicly available on the Data $\&$ Analysis Center for Exoplanets(DACE)\footnote{https://dace.unige.ch/radialVelocities/?pattern=HD$\%$2073344}. The spectra were reduced through the ELODIE Data
Reduction Software (TACOS), which also extracted the RVs
and activity indicators such as the full width at half maximum (FWHM) and the contrast of the cross-correlation function (CCF) through the cross-correlation technique \citep{Baranne1996}.

\subsection{Lick Fischer Radial Velocities}

Our analysis included 23 RVs of HD 73344 collected between January 1998 and January 2009 with the Hamilton Spectrograph \cite{Vogt1987} installed on the 3 meter Shane telescope on Mount Hamilton. The Hamilton Spectrograph has a wavelength range from 340 to 900 nm.  The resolution varies from 50,000 to 115,000 depending on the choice of slit width. The observations were part of the Lick Planet Search program \citep{Fischer2014} monitoring 387 bright FGKM dwarfs to search for  giant exoplanets.

\subsection{Automated Planet Finder  Radial Velocities}

Our analysis included 23 RVs of HD 73344 observed by Automated Planet Finder (APF) spectrograph between March 2018 and January 2019. APF  is a 2.4m robotic telescope at Lick Observatory, Mount Hamilton, designed to find and characterize exoplanets with high-cadence Doppler spectroscopy \citep{Radovan2014,Vogt2014}. The APF RVs are published in \cite{Rosenthal2021}. 

\subsection{SOPHIE  Radial Velocities}

Our analysis included 312 RVs of HD 73344 observed by at the echelle spectrograph SOPHIE at Haute-Provence
Observatory (OHP, France, \citealp{Perruchot2008}). The observations was taken from November 2018 to March 2020 using a high observing cadence of 3 points per night to average the stellar variability. The SOPHIE RVs are published in \cite{Sulis2024}. %In our analysis, we grouped these observations into one RV per night. The binned RV time series contain a total of 118 RVs.

\subsection{Keck/HIRES Radial Velocities}

Our analysis included 238 archival RVs taken with High Resolution Echelle Spectrometer (HIRES, \citealt{Vogt1994}) at the Keck I 10 m telescope on Maunakea between March 2018 and June 2021 from \cite{Sulis2024}. The archival HIRES RVs were collected with high cadence, up to three sets of five consecutive observations each night. This approach was designed to minimize sensitivity to stellar variations. In our analysis, we binned these observations into one RV per night, resulting in a total of 23 RVs.

Furthermore, we used 16 newly collected RVs of HD 73344 taken from 2021 to 2024 using Keck/HIRES. The observations are part of a survey aiming to search for substellar/stellar companions to transiting planet hosts with significant \textit{Hipparcos} and \textit{Gaia} astrometric accelerations \citep{zhang2023}. We used the standard California Planet Search (CPS) pipeline described in \cite{Howard2010} to determine RVs. Spectra were obtained with an iodine gas cell in the light path for wavelength calibration. An iodine-free template spectrum bracketed by observations of rapidly rotating B-type stars was used to deconvolve the stellar spectrum from the spectrograph PSF. We then forward-model the spectra taken with the iodine cell using the deconvolved template spectra \citep{Butler1996}. The wavelength scale, the instrumental profile, and the RV in each of the $\ttilde 700$ segments of 80 pixels were solved simultaneously \citep{Howard2010}. 

The RVs used in this work are presented in Table~\ref{tab:rvs}.
\begin{deluxetable}{cccc}

\tablecaption{HD 73344 RVs}\label{tab:rvs}
\tablehead{\colhead{Time} & \colhead{RV} & \colhead{$\sigma_{\mathrm{RV}}$} & \colhead{Inst} \\ 
\colhead{(BJD - 2450000)} & \colhead{(m/s)} & \colhead{(m/s)} & \colhead{} } 
\startdata
50770.7053 & 6231.95 & 9.94 & ELODIE\\
50858.4805 & 6247.14 & 8.97 & ELODIE\\
50890.4739 & 6154.99 & 8.89	& ELODIE\\
... & ... & ... &  ... \\
50831.8418 & 6.58 & 6.47 & Lick Fischer\\
50854.8232 & -62.43 & 9.59 & Lick Fischer\\
51173.9921 & -62.09 & 9.43 & Lick Fischer\\
... & ... & ... &  ... \\
58207.6360 & -53.42 & 5.68 & APF\\
58250.7309 & 19.30 & 4.64 & APF\\
58260.7041 & -12.07 & 4.63 & APF\\
... & ... & ... &  ... \\
58425.6567 & 6251.50 & 1.30 & SOPHIE\\
58426.6053 & 6255.80 & 2.10 & SOPHIE\\
58426.6955 & 6254.90 & 1.80 & SOPHIE\\
... & ... & ... &  ... \\
58194.8831 & -27.48 & 1.79 & HIRES\\
58194.8840 & -28.51 & 1.75 & HIRES\\
58194.88493 & -21.47 & 1.79 & HIRES\\
... & ... & ... &  ... \\
\enddata
\tablecomments{Times are in BJD - 2400000.0. The RV uncertainties do not include RV jitter. All RV data utilized in this paper, including those sourced from the literature, are available in a machine-readable format. }

\end{deluxetable}

\begin{table}
\setlength{\tabcolsep}{4pt}
\centering
\caption{Hipparcos-Gaia astrometric acceleration in declination and right ascension for HD 73344 from \cite{Brandt2021}. The $\Delta \mu_{\delta *}$ components have the $\cos{\delta}$ factor included. $\sigma[\Delta\mu]$ represent the uncertainties. We assume that the uncertainties on the proper motions and $\mu_{\rm{HG}}$ are independent and add them in quadrature to calculate these uncertainties (see \citealt{zhang2023} for details).}\label{tab:HGCA_obs}
\begin{tabular}{cccccc}
\hline
\hline
 Data & $\Delta\mu_{\alpha}$ & $\sigma[\Delta\mu_{\alpha}]$ & $\Delta\mu_{\delta *}$ & $\sigma[\Delta\mu_{\delta *}]$ & S/N\\
  epoch & \multicolumn{2}{c}{$\mathrm{mas}\ \mathrm{yr}^{-1}$} & \multicolumn{2}{c}{$\mathrm{mas}\ \mathrm{yr}^{-1}$} & \\
\hline
 {\it Gaia} & -0.109& 0.043 & 0.100 & 0.031 & 3.85 \\
 {\it Hipparcos} & 0.567  & 0.833 & -0.027  &  0.594 & 0.68 \\
\hline
\end{tabular}
\end{table}

\subsection{Hipparcos and Gaia astrometric acceleration\footnote{Also known as proper motion anomaly.}}\label{sec:HG}
We used the astrometric data for HD 73344 from the Hipparcos-Gaia Catalog of Accelerations (HGCA, \citealt{Brandt2021}). The HGCA catalog provides three proper motions in units of $\mathrm{mas}\ \mathrm{yr}^{-1}$: (1) the \textit{Hipparcos} proper motion $\mu_{\rm{H}}$ measured at an epoch near 1991.25; (2) the \textit{Gaia} EDR3 proper motion $\mu_{\rm{G}}$ measured at an epoch near 2016.01; (3) the long-term proper motion $\mu_{\rm{HG}}$ calculated as the difference in positions between \textit{Hipparcos} and \textit{Gaia} divided by the $\ttilde 25$-year baseline. Following \cite{Kervella2019} and \cite{Brandt2021}, we computed the astrometric acceleration by subtracting the long-term proper motion $\mu_{\rm{HG}}$ from the \textit{Hipparcos} or \textit{Gaia} proper motion. Table~\ref{tab:HGCA_obs} presents the astrometric accelerations of HD 73344 has a signal-to-noise ratio ($\rm{S/N}$) of 3.85 at the \textit{Gaia} epoch, indicating the existence of an unresolved companion. 

\subsection{TESS photometry}
HD 73344 was observed by the \textit{TESS} for Sectors 45, 46, 71 and 72 of the  mission, spanning from November 6 2021 to December 7, 2023. \textit{TESS} data were available for both 2-minute cadence and 20-second cadence light curves. The light curves were processed by the Science Processing Operations Center (SPOC) data reduction pipeline \citep{Jenkins2016}. 
We downloaded all four sectors of TESS Target Pixel File data and build Simple Aperture Photometry (SAP) light curves using the package Systematics-insensitive Periodogram (SIP, \citealt{Hedges2020}). 

%used a Systematics-insensitive Periodogram (SIP, \citealt{Hedges2020}) to download all five sectors of TESS Target Pixel File data and build Simple Aperture Photometry (SAP) light curves of the target.

%We downloaded all four sectors of Pre-search Data Conditioning simple aperture photometry (PDC$\_$SAP, \citealt{Smith2012, Stumpe2012,Stumpe2014})light curves, and then stitched and normalized them using the \textit{lightkurve} package \citep{LK2018}.

\newpage

\subsection{AO imaging}

\begin{figure}
    \centering
    \includegraphics[width=\linewidth]{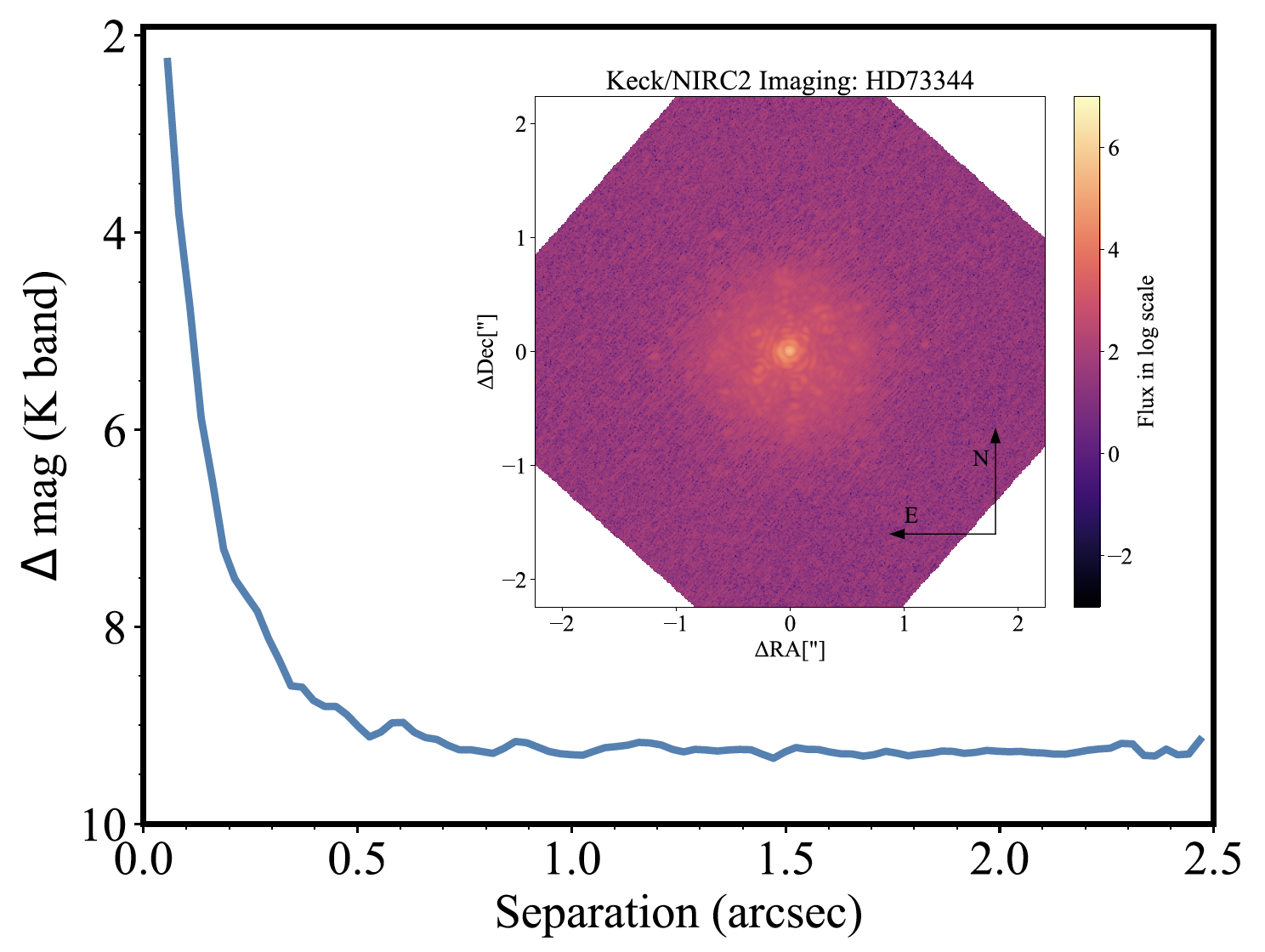}
    \caption{$5\sigma$ contrast curve from Keck/NIRC2 AO imaging of HD 73344. No stellar companions detected.}  
    \label{fig:figure_ao}
\end{figure}

We obtained Adaptive Optics (AO) imaging of HD 73344 to search for close stellar companions using the near-infrared imager (NIRC2) in the Kcont bandpass (2.2558 $\mu m$ -2.2854 $\mu m$) on the 10-meter Keck II telescope on January 02, 2023 UT. We used an exposure time of 0.2 second per coadd and 50 coadds per frame. We performed
flat-fielding, bad-pixel removal using the Vortex Imaging Processing (VIP) software package \citep{GandG1973, Christiaens2023}, and correct for geometric distortions by applying the solution in \cite{Service2016}. Since the data were taken in vertical angle mode, we de-rotated each image according to the parallactic angles and then stacked the individual images into a
combined image. To register the eight frames, we identify the position of the star by fitting a 2D Gaussian to the
stellar point spread function (PSF) in each frame. We then computed the $5\sigma$ contrast curves using package \texttt{PyKlip} \citep{Wang2015} by injecting fake companions into the dataset at various separations and position angles. Figure~\ref{fig:figure_ao} shows the
contrast curves and AO image. From the
contrast curve and AO image, we find that  HD 73344 has no stellar companion brighter than 11  magnitude in K band detected from $0.1^{\arcsec}$ to $0.5^{\arcsec}$. There is also no stellar companion brighter than 14  magnitude in K band
detected from $0.5^{\arcsec}$ to $2.5^{\arcsec}$. 

\section{Host Star Characterization}

\subsection{Stellar Properties} \label{subsec:starpara}
HD 73344 (EPIC 212178066, HIP 42403, TIC 175193677) is a bright F6V star with an age of $1.150^{+0.300}_{-0.326}$ Gyr \citep{Sulis2024} at a distance of $35.193^{+0.023}_{-0.014}\ \rm{pc}$ \citep{Gaia}. 
\cite{Sulis2024} measured the stellar mass as $1.20\pm{0.02}\ M_{\odot}$, stellar radius as $1.22\pm{0.04}\ R_{\odot}$ and [Fe/H] as $0.18\pm{0.043}$ using SOPHIE spectra. They also reported a spectroscopically determined projected rotation velocity, $v\sin i_*$ of $\ttilde 5.3\ \mathrm{km\ s^{-1}}$. 

%\citep{Petigura_thesis,Sulis2024}. 

Furthermore, we derived stellar properties based on the iodine-free spectra from Keck/HIRES using SpecMatch synthetic methodology (SpecMatch-Synth, \citealt{Petigura_thesis}). The stellar mass is $1.19\pm{0.05}\ M_{\odot}$, stellar radius is $1.22\pm{0.03}\ R_{\odot}$ and $v\sin i_*$ is $5.67\pm{1.0}~\rm{km}\ s^{-1}$. We adopt the HIRES-determined stellar properties for the star, which are listed and compared to literature values in Table~\ref{tab:star-compare}. 

\begin{deluxetable}{lcr}
\tablecaption{Stellar Parameters of HD 73344}
\label{tab:star-compare}
\tablehead{\colhead{Parameters (Unit) }&\colhead{This Work}  &\colhead{Reference}  }
\startdata
$T_{\rm eff}\ (\rm{K})$  & $\mathbf{6148\pm{100}}$    & $6220\pm{+64}$ (a)   \\
$M_{\star}$ ($\rm M_{\odot}$) & $\mathbf{1.19}\pm{0.05}$ & $1.20\pm{0.02}$ (a)   \\
$R_{\star}$ ($\rm R_{\odot}$) & $\mathbf{1.22\pm{0.03}}$ & $1.22\pm{0.04}$ (a) \\
log \textit{g} (cgs) & $4.36\pm{0.10}$ & $4.39\pm{0.02}$(a)   \\
$\rm [Fe/H]$ (dex)   &$0.16\pm0.06$ & $0.18\pm0.043$(a)    \\
$V_{\rm{mag}}$&  --  &$8.135\pm0.03$ (a)    \\
Age (Gyr)& --  & $1.150^{+0.300}_{-0.326}$   (a)   \\
$v\sin i_*$ ($\rm km\ s^{-1}$) & $5.67\pm{1}$ &  $\sim5.3$   (a)  \\
$\varpi$ (mas)&--  & $28.375\pm{0.021}$   (b)    \\
$\mathrm{P_{rot}}$ (days)   & $9.03\pm{1.40}$ & $9.09\pm{0.04}$ (a)    \\
$i_*$ (deg)&   $63.6_{-16.5}^{+17.4}$ &  $\sim 53$ (a)   \\
$k_2$ &   -- &  0.0034 (c)   \\
$C$ ($\rm M_* R_{*}^{2}$) &   -- &  0.056 (c)   \\
\enddata
\tablecomments{ (a). \citet{Sulis2024}; (b). \citet{Gaia}. (c). \citet{landin2009}. $k_2$ is the stellar second fluid Love number and $C$ is the stellar moment of inertia along the short axis (details see \citealt{BF14}). } 
\end{deluxetable}

%* $\mathrm{P_{rot}}$ is measured from TESS light curves described in Section~\ref{sec:istar}. $i_{*}$ is constrained with spectroscopic $v\sin i_*$ and stellar rotation period described in Section~\ref{sec:istar}.

%\citep{FP18,Sulis2024}.

\subsection{Stellar Rotation Period and stellar inclination}\label{sec:sr_si}

\begin{figure}
    \centering
    \includegraphics[width=\linewidth]{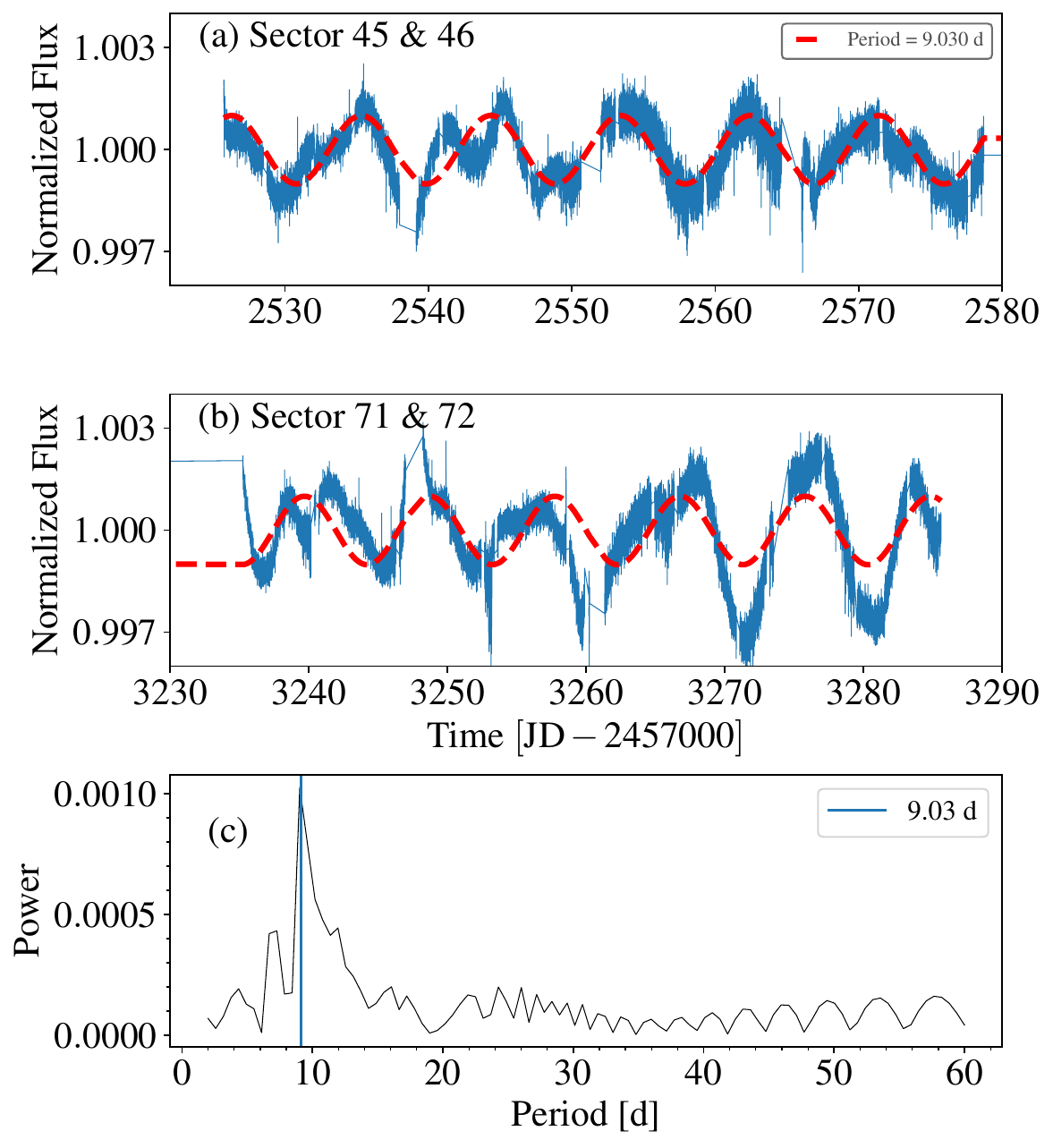}
    \caption{Panel a-b: TESS light curves of HD 73344 obtained in Sector 45, 46, 71 and 72 with the modulation model at a period of 9.03 days. Panel c: the Lomb-Scargle periodogram of the detrended TESS lightcurves of HD 73344 using SIP \citep{Hedges2020}. The highest peak in the periodogram is located at 9.03 days. }
    \label{fig:figurerot}
\end{figure}

Figure~\ref{fig:figurerot} presents the 2-min cadence TESS SPOC light curves from sectors 45, 46, 71 and 72 of HD 73344 and the periodogram of the stitched light curves. We stitched and removed systematics in the light curves using open-source package Systematics-insensitive Periodogram (SIP, \citealt{Hedges2020}). We determined a rotation period $P_\mathrm{rot}$ of 9.03 days. Our result is consistent with the values reported by  \cite{Sulis2024}, who utilized light curves from \textit{K2}, \textit{TESS}, and radial velocity (RV) data. We obtained an uncertainty of the rotation period as 1.3 days, from the width of the periodogram peak and an extra $10\%$ to account for  surface differential rotation \citep{Epstein2014, Claytor2022}.

\begin{figure}
    \centering
    \includegraphics[width=0.8\linewidth]{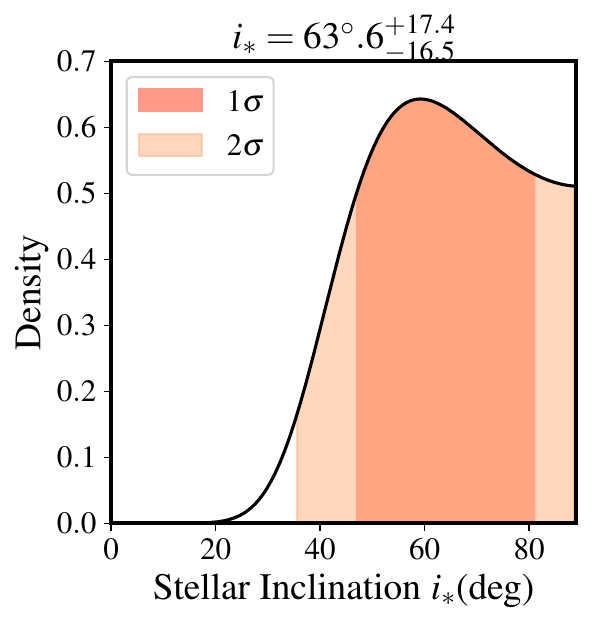}
    \caption{ Posterior distribution of stellar inclination. The dark and light shadow present the $1\sigma$ and $2\sigma$ credible region, respectively.}  
    \label{fig:figurestellar_inc}
\end{figure}

With the measurement of stellar rotation period $P_{rot}$, we can derive the the equatorial rotational velocities of the star as $ v_{eq} = 2\pi R_{*}/P_{\rm{rot}}$, where $R_*$ is the stellar radius. In addition, the spectroscopically determined projected rotation velocity, $v\sin i_*$, is $5.67\pm{1.0}\ \mathrm{km\ s^{-1}}$ \citep{Petigura_thesis}. Because  $v_{eq} $ and $v\sin{i_*}$ are not statistically independent \citep{MandW2020M}, we can not simply derive the stellar inclination as $\sin^{-1}(\frac{ v\sin{i_*}}{v_{eq}})$. To consider the correlation between $v_{eq} $ and $v\sin{i_*}$, we adopted the Bayesian probabilistic framework from \cite{Bowler2023}. Figure~\ref{fig:figurestellar_inc}  shows the resulting distribution of the stellar inclination, which is away from $90^{\circ}$ with $1 \sigma$ significance. We obtained a stellar inclination  peaking at $58^{\circ}$ and the median value along with $1\sigma$ uncertainties is $i_{\star}=63^{\circ}.6^{+17.4}_{-16.5}$.

%\begin{equation}{\label{eqn:i}}
%P(i_*\mid P_\mathrm{rot}, R_*, v\sin i_*) \propto  \sin i_* \times \frac{e^{- \frac{\big(v \sin i_* - \frac{2\pi R_*}{P_\mathrm{rot}}\sin i_* \big)^2}{2\big(\sigma_{v\sin i_*}^2 + \sigma_{v_\mathrm{eq}}^2 \sin^2 i_* \big)}} }{\sqrt{\sigma_{v\sin i_*}^2 + \sigma_{v_\mathrm{eq}}^2 \sin^2 i_*}},
%\end{equation}
%where $\sigma_{v\sin i_*}$ and $\sigma_{v_\mathrm{eq}}$ are the uncertainties in the projected velocity and the equatorial velocity. 

\section{The discovery of an outer giant planet}

\subsection{Periodicity Analysis}

\begin{figure}
    \centering
    \includegraphics[width=\linewidth]{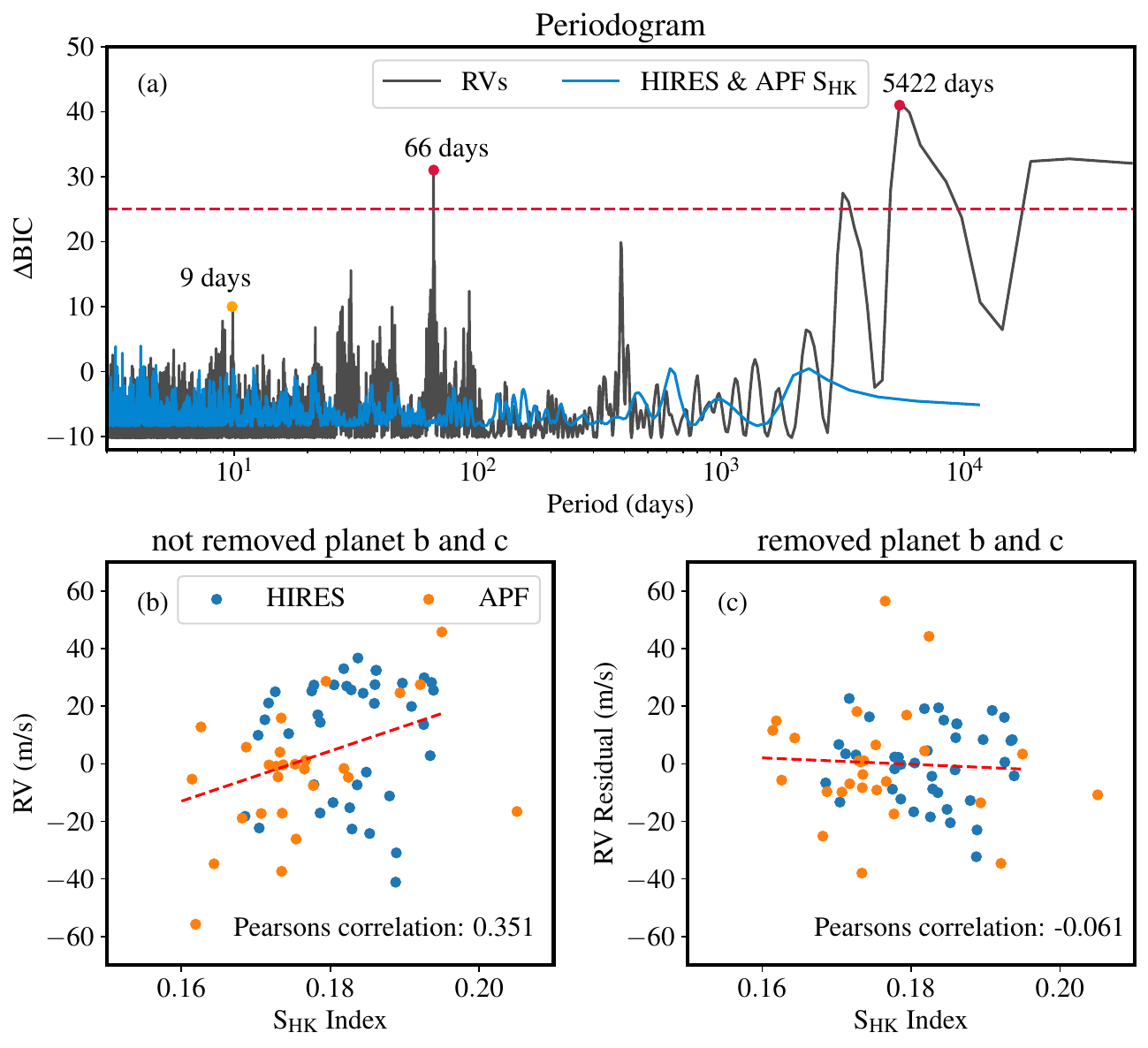}
    \caption{(a) Black line shows the periodogram of all RVs (except those from SOPHIE) computed using \textit{RVsearch}. Blue line shows the periodogram of $\rm{S_{HK}}$ index from HIRES and APF observations. The red dashed line corresponds to $0.1\%$ false alarm probability. (b) HIRES and APF RVs vs. $\rm{S_{HK}}$ index. (c) HIRES and APF RV residuls after removing signal from HD 73344 b $\&$ c vs. $\rm{S_{HK}}$ index. }  
    \label{fig:figure1}
\end{figure}

Figure~\ref{fig:figure1}(a) shows the periodogram of over 27 years of radial velocity data for HD 73344, computed using the \textit{RVsearch} package \citep{Rosenthal2021}. We excluded the SOPHIE RVs from the periodogram analysis because these observations were obtained at high cadence, with three data points per night. The high-cadence SOPHIE RVs were primarily designed to mitigate stellar noise and characterize the orbital parameters of the transiting planet with a 15.61-day period \citep{Sulis2024}. Including these high-cadence RVs could potentially obscure signals at longer periods. However, these SOPHIE RVs were included in the orbital fitting analysis.

Specifically, we used the \textit{RVsearch} package \citep{Rosenthal2021} to  establish an orbital period grid ranging from two days to five times the baseline of observational data to search for periodic signals. The algorithm  iteratively fits a sinusoid at each period and calculates the Bayesian Information Criterion (BIC) difference, $\Delta \rm{BIC}$. It also estimates the false alarm probability (FAP) by fitting a linear model to a log-scale histogram of periodogram power values (see \citealt{Howard2016} for details). The detection threshold of  $0.1\%$ FAP  means that only $0.1\%$ of periodogram values are above this threshold. 

The most significant peak surpassing the detection threshold in the periodogram is at 5422 days. There is also a smaller peak around 3000 days, which represents the harmonics of the signal observed at 5422 days. We also observed a pronounced peak at 66 days.  \cite{Sulis2024} previously reported the peak at 66 days and attributed it from a non-transiting planet. Similar to \cite{Sulis2024}, we did not detect the 15.6-day period signal of the transiting planet HD 73344 b in the RV periodogram. Their analysis has shown that the short-term stellar activity masking the planet's signature. Additionally, we observed a peak at $\ttilde 9$ days, which corresponds to the star's rotation period.

\subsection{Stellar Activity}

To investigate whether the long-period signal is from the stellar activity, we also analyzed the $\rm{S_{HK}}$ index from APF and HIRES spectra by measuring the core flux of Calcium H $\&$ K lines. The blue line in figure~\ref{fig:figure1} (a） presents the periodogram of  $\rm{S_{HK}}$ index. However, the $\rm{S_{HK}}$ data are only available from APF and HIRES, which provides a significantly shorter baseline. As a result, the periodogram of the $\rm{S_{HK}}$ index reveals no significant peaks beyond 3000 days.

Figure~\ref{fig:figure1} (b)$\&$(c) present $\rm{S_{HK}}$ index vs. the original RVs and RV residuals after removing the inner planets, respectively. In either case, the Pearson correlation coefficients indicate no strong  correlation between RVs and $\rm{S_{HK}}$ index. However, it is important to note that APF and HIRES data cover only a small portion of the planetary orbit. Therefore, a longer period of monitoring stellar activity is essential to further investigate the long-term effects of stellar activity.

On the other hand, the observed Hipparcos-Gaia astrometric acceleration of HD 73344 could offer complementary support for the existence of a giant planet, as the astrometric measurements are not influenced by stellar activity.

\subsection{RV-only Keplerian Fit}
\begin{figure}
    \centering
    \includegraphics[width=\linewidth]{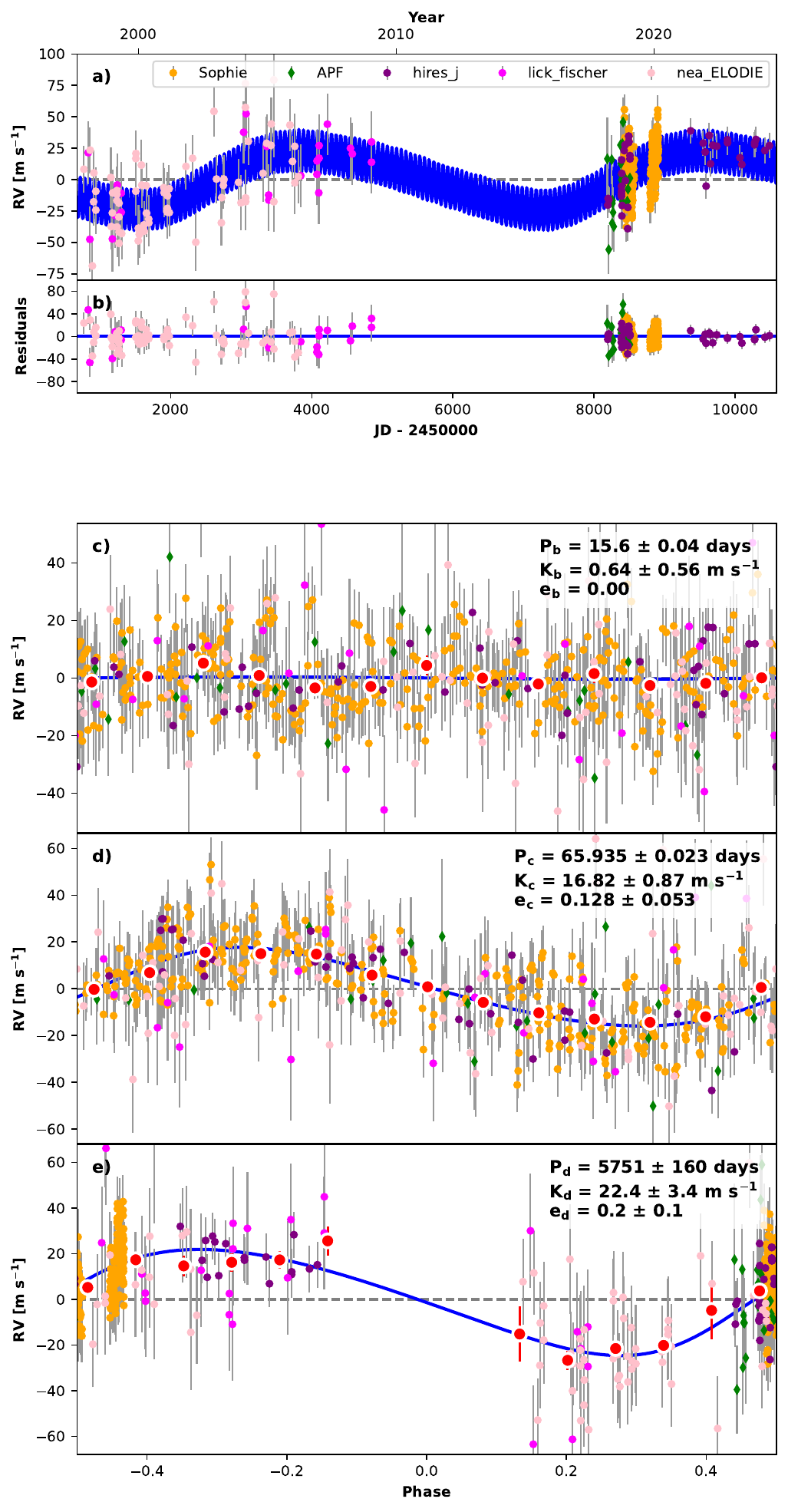}
    \caption{ Best-fit 3-planet Keplerian orbital model for HD 73344. a: HD 73344 RVs with errors (black) and their best fit model (blue) as a function of time. b: the residuals. c$\sim$e: RV data and models for each planet phase-folded at the best-fit orbital period with all other planets' signals removed.  }  
    \label{fig:figure2}
\end{figure}

\begin{deluxetable*}{lllll}
\tablecaption{ MCMC Posteriors for RV-only fitting}\label{tab:rvparams}
\tablehead{
  \colhead{Param.} & 
  \colhead{ Credible Interval} & 
  \colhead{Max Likelihood} & 
  \colhead{Units} & 
  \colhead{Prior$^{*}$}
}
\startdata
\multicolumn{4}{c}{\bf{Fitted parameters}} \\[1pt]
\hline
  $P_{b}$ & $15.604^{+0.044}_{-0.036}$ & $15.6$ & days & $\mathcal{N}(15.611 ,0.05)$  \\
  $T\rm{conj}_{b}$ & $2458484.6^{+1.1}_{-1.0}$ & $2458484.7$ & JD & $\mathcal{N}(2458484.901 ,1)$ \\
  $e_{b}$ & $\equiv0.0$ & $\equiv0.0$ & & $\mathcal{U}(0 ,0.99)$ \\
  $\omega_{b}^{\dagger}$ & $\equiv0.0$ & $\equiv0.0$ &  radians & $\mathcal{U}(0 ,2\pi)$\\
  $K_{b}$ & $<2.82\ (3\sigma)$ & $0.44$ & m s$^{-1}$ & $\mathcal{U}(0 ,\infty)$ \\
  $P_{c}$ & $65.936^{+0.023}_{-0.022}$ & $65.936$ & days  & $\mathcal{N}(65.92 ,1)$ \\
  $T\rm{conj}_{c}$ & $2455482.98^{+0.92}_{-0.89}$ & $2455483.01$ & JD & $\mathcal{N}(2455482.94 ,1)$ \\
  $e_{c}$ & $0.124^{+0.052}_{-0.053}$ & $0.128$ & & $\mathcal{U}(0 ,0.99)$ \\
  $\omega_{c}^{\dagger}$ & $1.96^{+0.43}_{-0.41}$ & $-1.18$ & radians  & $\mathcal{U}(0 ,2\pi)$\\
  $K_{c}$ & $16.83\pm 0.87$ & $16.8$ & m s$^{-1}$ & $\mathcal{U}(0 ,\infty)$ \\
  $P_{d}$ & $5746^{+170}_{-150}$ & $5699$ & days & $\mathcal{N}(5673 ,1000)$ \\
  $T\rm{conj}_{d}$ & $2444345^{+100}_{-550}$ & $2444325$ & JD &-- \\
  $e_{d}$ & $0.2\pm{0.1}$ & $0.175$ & & $\mathcal{U}(0 ,0.99)$ \\
  $\omega_{d}^{\dagger}$ & $1.07^{+0.62}_{-0.58}$ & $-1.99$ & radians & $\mathcal{U}(0 ,2\pi)$ \\
  $K_{d}$ & $22.7\pm 3.4$ & $23.4$ & m s$^{-1}$ & $\mathcal{U}(0 ,\infty)$ \\
\hline
\multicolumn{4}{c}{\bf{Other parameters}} \\[1pt]
\hline
  $\gamma_{\rm ELODIE}$ & $\equiv6223.9168$ & $\equiv6223.9168$ & m s$-1$ &-- \\
  $\gamma_{\rm Lick\ Fischer}$ & $\equiv-14.7772$ & $\equiv-14.7772$ & m s$-1$ &-- \\
  $\gamma_{\rm HIRES}$ & $\equiv-2.0331$ & $\equiv-2.0331$ & m s$-1$ &-- \\
  $\gamma_{\rm APF}$ & $\equiv0.0938$ & $\equiv0.0938$ & m s$-1$&-- \\
  $\gamma_{\rm SOPHIE}$ & $\equiv6235.7652$ & $\equiv6235.7652$ & m s$-1$&-- \\
  $\dot{\gamma}$ & $\equiv0.0$ & $\equiv0.0$ & m s$^{-1}$ d$^{-1}$ &--\\
  $\ddot{\gamma}$ & $\equiv0.0$ & $\equiv0.0$ & m s$^{-1}$ d$^{-2}$ &--\\
  $\sigma_{\rm ELODIE}$ & $16.7^{+2.6}_{-2.2}$ & $16.0$ & $\rm m\ s^{-1}$ &$\mathcal{U}(0 ,50)$\\
  $\sigma_{\rm Lick\ Fischer}$ & $25.3^{+5.3}_{-4.0}$ & $24.0$ & $\rm m\ s^{-1}$  &$\mathcal{U}(0 ,50)$\\
  $\sigma_{\rm HIRES}$ & $9.65^{+0.26}_{-0.48}$ & $10.0$ & $\rm m\ s^{-1}$  &$\mathcal{U}(0 ,10)$\\
  $\sigma_{\rm APF}$ & $19.8^{+3.8}_{-2.8}$ & $19.1$ & $\rm m\ s^{-1}$  &$\mathcal{U}(0 ,50)$\\
  $\sigma_{\rm SOPHIE}$ & $11.76^{+0.51}_{-0.5}$ & $11.63$ & $\rm m\ s^{-1}$  &$\mathcal{U}(0 ,50)$\\
\hline
\multicolumn{4}{c}{\bf{Derived parameters}} \\[1pt]
\hline
  $M_b$ & $<12.43\ (3\sigma)$ & $0.5$ & M$_{\oplus}$ &--\\
  $a_b$ & $0.1296^{+0.0016}_{-0.0017}$ & $0.1325$ &  AU &--\\
  $M_c\sin i$ & $0.373^{+0.022}_{-0.021}$ & $0.373$ & M$_{\rm Jup}$ &--\\
  $a_c$ & $0.3388^{+0.0042}_{-0.0043}$ & $0.3463$ &  AU &--\\
  $M_d\sin i$ & $2.21^{+0.34}_{-0.35}$ & $2.03$ & M$_{\rm Jup}$ &--\\
  $a_d$ & $6.66^{+0.16}_{-0.15}$ & $6.9$ &  AU &-- \\
\hline
\enddata
\tablenotetext{}{$*$ $\mathcal{U}$ stands for a uniform distribution, with two numbers representing the lower and upper boundaries. $\mathcal{N}$ stands for a Gaussian distribution, with numbers representing the mean and standard deviation.}
\tablenotetext{}{$\dagger$ The default output from RadVel is the argument of periastron for the star. Here, we report the argument of periastron for the planets by adding an offset of $\pi$. }
\label{tab:params}
\end{deluxetable*}

We first performed a three-planet Keplerian orbital fitting to RVs using \textit{RadVel} package \citep{Fulton2018}. The model includes five orbital elements for each planet $K$, $P$, $T_{conj}$, $\sqrt{e}sin\omega$,$\sqrt{e}cos\omega$, where $K$ is the RV semi-amplitude, $P$ is the orbital period, $T_{conj}$ is the time of conjunction, $e$ is the eccentricity, and $\omega$ is the argument of pericenter. For HD 73344 b, we set the eccentricity and argument of pericenter as 0 to simplify the fitting because small transiting planets are likely to have low eccentricities \citep{Van2015}. We also used Gaussian priors informed by \textit{K2} and \textit{TESS} observations \citep{Yu2018, Sulis2024} for the orbital period $P_b$ and conjunction time $T_{conj,b}$ of HD 73344 b. 
For the other two planets HD 73344 c $\&$ d, we allow all five of its orbital parameters to vary. We set bounds on $0<e<1$, $K>0$ for all planets. We set bounds on the jitter terms as $0-10\ \mathrm{m\ ^{-1}}$ for HIRES and $0-50\ \mathrm{m\ ^{-1}}$ for other instruments. 

We performed the Markov Chain Monte Carlo (MCMC) exploration with \textit{emcee} \citep{FM2013} to estimate parameter credible intervals.  Our MCMC analysis used 50 walkers and ran for $1.5 \times 10^6$ steps per walker, achieving a maximum Gelman-Rubin(GR) statistic of 1.004. Figure~\ref{fig:figure2} shows the best fit Keplerian solution. Similar to \cite{Sulis2024}, we obtained an upper limit for the mass of HD 73344 b as $M_b<12.43\ M_{\oplus}$ at $3 \sigma$. We constrained the minimum mass of HD 73344 c to be $m_{c}\sin{i_c}=0.373^{+0.022}_{-0.021} \ \rm M_{J}$, which is consistent with \cite{Sulis2024} within $1\sigma$. Furthermore, we derived a minimum mass for HD 73344 d from RV amplitudes as $m_{d}\sin{i_d}=2.21^{+0.34}_{-0.35} \ \rm M_{J}$, semi-major axis as $a_d = 6.66^{+0.16}_{-0.15}$ AU and an eccentricity of $e_d=0.2\pm{0.1}$. The derived planetary parameters are given in Table~\ref{tab:rvparams}.

\subsection{Three-dimensional orbit fitting of HD 73344 d}\label{sec:3dfit}

\begin{figure*}
    \centering
    \includegraphics[width=\linewidth]{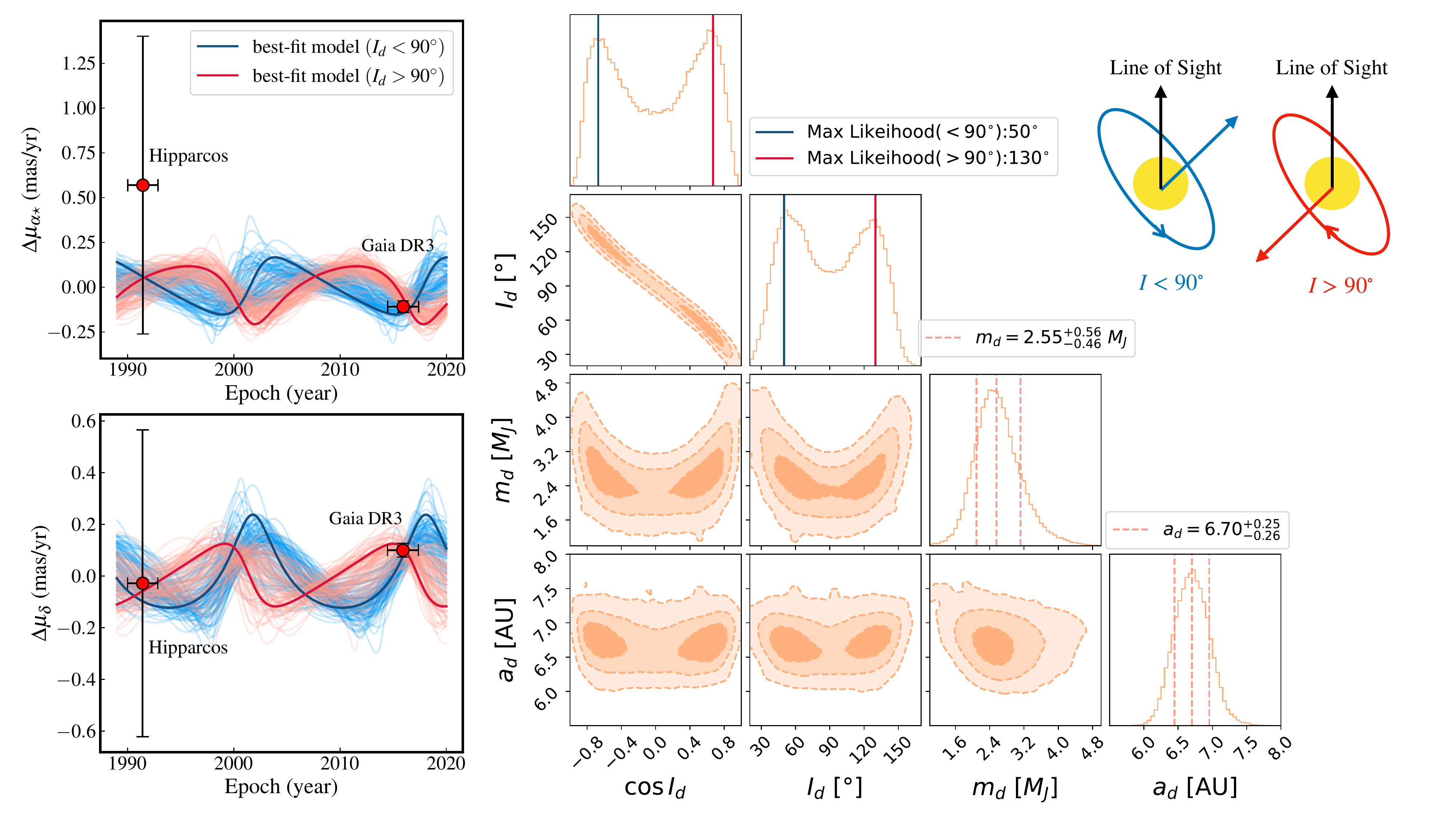}
    \caption{ Left: observed and fitted Hipparcos and Gaia proper motions of HD 73344 in right ascension (top) and declination (bottom). The blue and red lines are models with orbital inclination smaller or larger than $90^{\circ}$. Right: The joint posterior distributions for cosine term of orbital inclination $\cos{I_d}$, orbital inclination $I_d$, mass $m_d$, and semi-major axis $a_d$ of HD 73344 d from combined fitting of RV and Hipparcos-Gaia astrometric acceleration. Moving outward, the dashed lines on the corner plots correspond to $1\sigma$, $2\sigma$ and $3\sigma$ contours. For $\cos{I_d}$ and $I_d$, red and blues lines correspond to two peaks symmetrically positioned about 90 degrees, which arise from the degeneracy between prograde and retrograde orbits. The orange dashed lines in the 1D distribution of mass $m_d$ and semi-major axis $a_d$ denote median values with $1\sigma $ interval. The cartoon in the upper right corner shows the prograde and retrograde orbits for a planet in the same orbital plane. The black arrows show the line of sight direction, and blue and red arrows are the orbital axis of the planet. }  
    \label{fig:figure5}
\end{figure*}
\begin{figure*}
    \centering
    \includegraphics[width=0.9\linewidth]{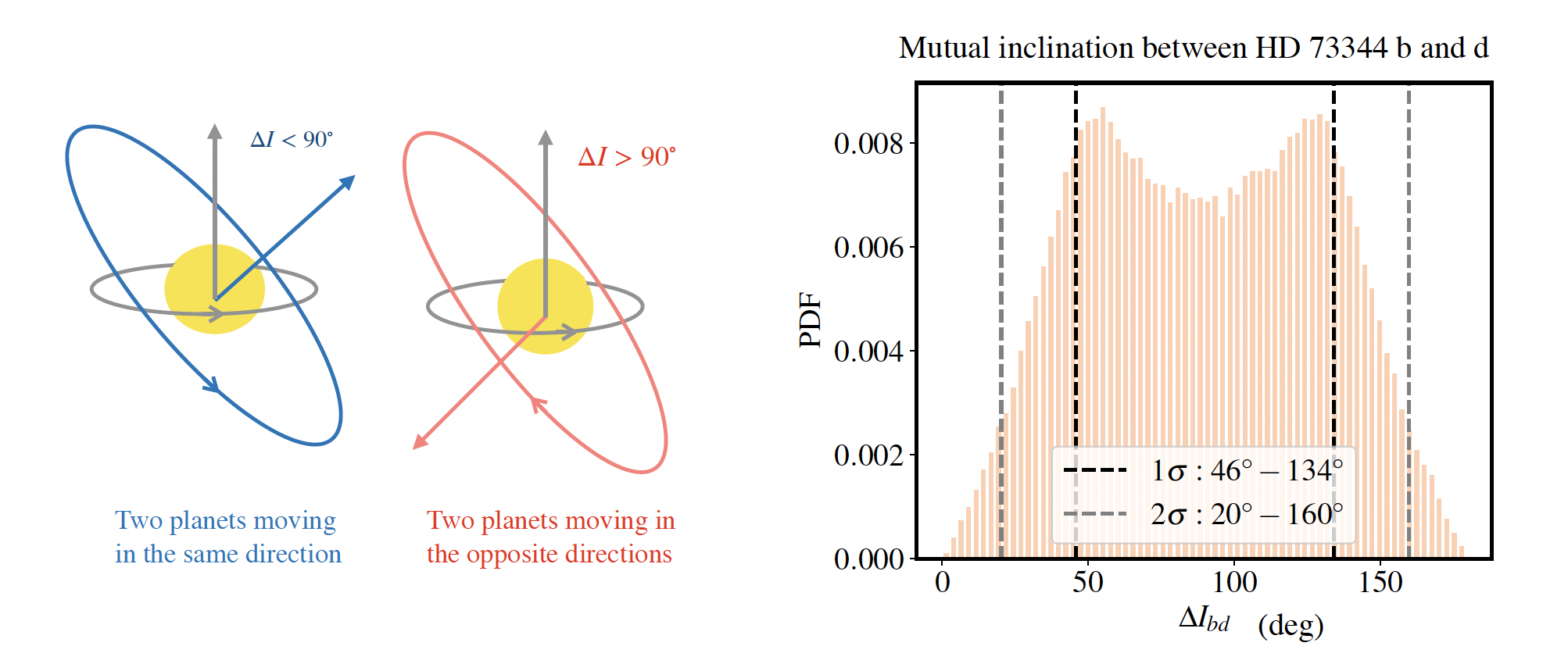}
    \caption{ Left: Sketch to show the mutual inclinations between two planets orbiting in the same direction ($\Delta I <90^{\circ}$) and orbiting in opposite directions ($\Delta I >90^{\circ}$), respectively. The gray circle and arrow show the orbit and orbital axis of the inner planet. The blue and red circle and arrows denote the orbits and orbital axes of the an outer planet in two scenarios. Right: Posterior distribution of mutual inclination between HD 73344 b and d. }  
    \label{fig:figure6}
\end{figure*}
We then performed a joint fit to the RVs and Hipparcos-Gaia astrometric accelerations to characterize the three-dimensional (3D) orbit of the outer giant planet. The long-baseline RVs over 27 years offer crucial constraints on the orbit of the outer giant planet, including orbital period and eccentricity, while the two astrometric accelerations with a baseline around 25 years help to break the degeneracy between the planet's mass and orbital inclination. Our model includes seven parameters to describe the orbits and mass of the outer giant planet HD 73344 d: the true mass of planet $m_{d}$, the cosine  of its orbital inclination $\cos{I_{d}}$, the orbital period $P_{d}$, the longitude of the ascending node $\Omega_{d}$, the eccentricity $e_{d}$, and the argument of periastron $\omega_{d}$ (fitted as $\sqrt{e_{d}}\cos{\omega_{d}}$ and $\sqrt{e_{d}}\sin{\omega_{d}}$)\footnote{Note that $\omega_{d}$ is the argument of periastron of the {\it planet}, not the star. The argument of periastron of the planet is offset by $\pi$ from that of the star.}, and the epoch of periastron $\tau_{d}$ at a reference epoch ($t_{\rm{ref}}=2454574.5$ JD). $\tau$ is a dimensionless quantity ranging from 0 to 1, calculated relative to a reference epoch $t_{\mathrm{ref}}$, the time of pariastron $t_{\mathrm{p}}$ and the planet orbital period $P$ : $\tau = (t_{p}-t_{\mathrm{ref}})/P$ \citep{Blunt2020}. We have another five parameters for the orbit of HD 73344 c: RV semi-amplitude $K_{c}$, orbital period $P_{c}$, $\sqrt{e_{c}}\cos{\omega_{c}}$, $\sqrt{e_{c}}\sin{\omega_{c}}$ and mean longitude at a reference epoch $\tau_{c} $. Finally, we included two parameters for the  mass of the host star $M_{\star}$ and the parallax of the system $\varpi$, and ten parameters to account for the RV zero points ($\gamma_{\rm{SOPHIE}}$,$\gamma_{\rm{ELODIE}}$, $\gamma_{\rm{Lick\_Fischer}}$, $\gamma_{\rm{APF}}$, $\gamma_{\rm{HIRES}}$) and jitter terms ($\sigma_{\rm{SOPHIE}}$, $\sigma_{\rm{ELODIE}}$, $\sigma_{\rm{Lick\_Fischer}}$, $\sigma_{\rm{APF}}$, $\sigma_{\rm{HIRES}}$)of all instruments.  

\begin{deluxetable*}{lcc}
\tablecaption{MCMC Posteriors for 3-D orbital fitting}\label{tab:3d}
\setlength{\tabcolsep}{0.10in}
\tablewidth{0pt}
\tablehead{
\colhead{Parameter (Unit)}              &
\colhead{Median $\pm$1$\sigma$} &
\colhead{Prior$^{4}$}                 }
\startdata
\multicolumn{3}{c}{Fitted parameters} \\[1pt]
\hline
\multicolumn{3}{c}{} \\[-5pt]
\multicolumn{3}{l}{\textbf{HD 73344~c}}\\[3pt]
RV semi-amplitude of 73344~c $K_{c}$ ($\mathrm{m\,s^{-1}}$)                                      & $16.36^{+1.15}_{-1.12}$                      & $\mathcal{N}(16 ,5)$                                                 \\[3pt]
Orbital periods $P_{c}$ (days)                                                     & $65.94\pm{0.02}$                & $\mathcal{N}(66 , 10)$                                                   \\[3pt]
Eccentricity term$\sqrt{e_c}\cos{\omega_{c}}$                                                      & $0.06^{+0.18}_{-0.20}$            & $\mathcal{U}(-1 , 1)$                                                            \\[3pt]
Eccentricity term$\sqrt{e_{c}}\sin{\omega_{c}}$                                                      & $-0.24^{+0.20}_{-0.12}$              & $\mathcal{U}(-1 , 1)$                                                             \\[3pt]
Epoch of periastron at 2454574.5 JD,  $\tau_{\rm c}$ & $0.35^{+0.15}_{-0.14}$                         & $\mathcal{U}(0 , 1)$        \\[3pt]
\multicolumn{3}{l}{\textbf{HD 73344~d}}\\[3pt]
Planet mass $m_{\rm d}$ ($\mathrm{M_{J}}$)                                       & $2.55^{+0.56}_{-0.46}$                           & $\mathcal{U}(0 , 300)$                                                  \\[3pt]
Orbital periods $P_{d}$ (days)                                                     & $5823_{-308}^{+312}$                &  $\mathcal{U}(0 , \infty)$                                                      \\[3pt]
Cosine term of inclination $\cos{I_{d}}^{1}$                                                   & $\mathbf{0.52_{-0.30}^{+0.22}}$              &  $\mathcal{U}(-1 , 1)$                              \\[3pt]
Longitude of ascending node  $\Omega_{d}$  (deg)                                &  $144_{+55}^{-43}$                &  $\mathcal{U}(0 , 360)$                                                             \\[3pt]
Eccentricity term$\sqrt{e}_{d}\cos{\omega_{d}}$                                                    & $-0.20^{+0.30}_{-0.27}$               &  $\mathcal{U}(-1 , 1)$                           \\[3pt]
Eccentricity term$\sqrt{e_{d}}\sin{\omega_{d}}$                                          
            &  $-0.27^{+0.34}_{-0.20}$              & $\mathcal{U}(-1 , 1)$                                                              \\[3pt]
Epoch of periastron at 2454574.5 JD, $\tau_{\rm c}$ & $0.59^{+0.12}_{-0.17}$                       & $\mathcal{U}(0 , 1)$          \\[3pt]
\multicolumn{3}{l}{\textbf{Others}}\\[3pt]
Host-star mass $M_{*}$ ($\mathrm{M_{\odot}}$)                                       & $1.18\pm0.05$                      & $\mathcal{N}(1.18 ,0.09)$                                                    \\[3pt]
Parallax $\varpi$ (mas)                                                              & $23.38\pm0.02$                          & $\mathcal{N}(\varpi_{\rm DR3} ,\sigma[\varpi_{\rm DR3}])^{2}$ \\[3pt]
RV zero point $\gamma_{\rm{SOPHIE}}$($\mathrm{m\,s^{-1}}$ )                                      &$6236.93^{+5.09}_{-4.91}$                     & --                                                            \\[3pt]
RV zero point $\gamma_{\rm{ELODIE}}$($\mathrm{m\,s^{-1}}$ )                                      &$6221.95^{+4.55}_{-4.73}$                     & --                                                            \\[3pt]
RV zero point $\gamma_{\rm{HIRES}}$($\mathrm{m\,s^{-1}}$ )                                      &       $-0.98^{+4.56}_{-3.95}$                       & --                                                             \\[3pt]
RV zero point $\gamma_{\rm{APF}}$($\mathrm{m\,s^{-1}}$ )                                      &$-0.20^{+5.88}_{-5.15}$                     & --                                                             \\[3pt]
RV zero point $\gamma_{\rm{lick\ fischer}}$($\mathrm{m\,s^{-1}}$ )                                      &       $-13.52^{+5.46}_{-5.32}$                       & --                                                            \\[3pt]
RV jitter term $\sigma_{\rm{SOPHIE}}$($\mathrm{m\,s^{-1}}$)                                      &     $11.81^{+0.76}_{-0.70}$                       & $\mathcal{U}(0,30)$                                                           \\[3pt]
RV jitter term $\sigma_{\rm{ELODIE}}$($\mathrm{m\,s^{-1}}$)                                      &     $16.49^{+2.32}_{-2.15}$                       & $\mathcal{U}(0,30)$                                                           \\[3pt]
RV jitter term $\sigma_{\rm{HIRES}}$($\mathrm{m\,s^{-1}}$)                                      & $11.96^{+1.62}_{-1.33}$                            & $\mathcal{U}(0,30)$                                                           \\[3pt]
RV jitter term $\sigma_{\rm{APF}}$($\mathrm{m\,s^{-1}}$)                                      &     $18.08^{+2.91}_{-2.37}$                       & $\mathcal{U}(0,30)$                                                           \\[3pt]
RV jitter term $\sigma_{\rm{lick\ fisher}}$($\mathrm{m\,s^{-1}}$)                                      & $23.69^{+3.48}_{-3.33}$                            & $\mathcal{U}(0,30)$                                                           \\[3pt]
\hline
\multicolumn{3}{c}{} \\[-5pt]
\multicolumn{3}{c}{Derived parameters }\\[1pt]
\hline
\multicolumn{3}{c}{} \\[-5pt]
$\rm{Inclination}$ $I_{d}^{1}$ (deg)                                   & $\mathbf{59_{-17}^{+19}}$                & --                                                            \\[3pt]
Semi-major axis $a_{d}$ (AU)                                                     & $6.70_{-0.26}^{+0.25}$                             & --                                                           \\[3pt]
Eccentricity $e_{d}$                                                            & $0.18^{+0.14}_{-0.12}$                 & --                                                          \\[3pt]
Argument of periastron $\omega_{d}$ (deg$)^{3}$                                   & $58.9_{-79.0}^{+57.3}$                & --                                                            \\[3pt]
Time of periastron $T_{p,c}$ (JD) & $2458061_{-1148}^{+735}$    & --                                                            \\[3pt]
\enddata
\tablecomments{   1. There are double peaks in the distribution of inclination due to the pro-grade and retro-grade degeneracy. We only present the solutions with $I_{d}>90^{\circ}$ here. 2.$\varpi_{\rm DR3}$ and $\sigma[\varpi_{\rm DR3}]$ present the parallax and parallax uncertainty of HD 118203 from Gaia DR3 observations. 3. We report the periastron of argument for planets here. 4. $\mathcal{U}$ stands for a uniform distribution, with two numbers representing the lower and upper boundaries. $\mathcal{N}$ stands for a Gaussian distribution, with numbers representing the mean and standard deviation.} %5. The HIRES RVs are median-subtracted, whereas the ELODIE RVs are not.1.The $\chi^{2}$ of RVs is 77 for 80 measurements. }
\end{deluxetable*}

We used the RV and astrometric models described in \cite{Xuan2020}.Given that the orbital periods of HD 73344 b and c (15 days and 66 days) are significantly shorter than the observation periods of both \textit{Hipparcos} ($\delta_{\rm{H}}=1227$ days) and \textit{Gaia} ($\delta_{\rm{GDR3}}=1002$ days), their astrometric signals are expected to be undetectable due to the small displacement of the star and also the smearing effect \citep{Kervella2019}. Therefore, we did not consider the inner planets in the astrometric model and did not fit their inclinations. We only consider a single-planet model (planet d only) to fit the astrometric data. Additionally, in our model,  we computed the tangential velocities at every individual observation time \footnote{ The \textit{Hipparcos} observation times can be found from the Hipparcos Epoch Photometry Annex \citep{vanLeeuwen1997} and the \textit{Gaia} DR3 observation times can be downloaded in \textit{Gaia} Observation Forecast Tool https://gaia.esac.esa.int/gost/} within the {\it Gaia} DR3 and {\it Hipparcos} observing periods, and then average over them to mimic the smearing effect. In addition, we do not consider HD 73344 b in our RV model because the RV semi-amplitude from HD 73344 b ($<2.82 m^{-1}$) is comparable or smaller than the RV jitter values. Therefore, we  consider a two-planet model (planet c and d) to fit the RV data.

We used the parallel-tempering Markov chain Monte Carlo (PT-MCMC) ensemble sampler in \textit{emcee} \citep{FM2013} to sample the parameter space with 40 different temperatures and 100 walkers.  Our results are taken from the `coldest' chain, which corresponds to the original, unmodified likelihood function.  Our PT-MCMC analysis  stabilized in the mean and root mean square (rms) of the posteriors of each of the model parameters after $1.2\times 10^{5}$ steps. We saved every 100th step of our chains and discarded the first $30\%$ of the chain as the burn-in portion.

Figure~\ref{fig:figure5} presents the astrometric data and models, and the joint posterior distributions for
the orbital inclination, mass and semi-mahor axis for HD 73344 d (see Appendix~\ref{fig:figureA2} for additional parameters). We obtained a bimodal distribution in the orbital inclination, symmetric around $90^{\circ}$. Additionally, the sum of values at two peaks equals to $180^{\circ}$. These two peaks result from the prograde-retrograde degeneracy, indicating that the planet orbits in the same plane but can move in two opposite directions. Due to only having astrometric data from two epochs and significant uncertainties in the \textit{Hipparcos} data, we are unable to resolve this degeneracy. The future release of time-series asymmetric data from Gaia DR4 will provide a better constraint on $I_d$ and help to break the degeneracy. If we only consider solutions with $I_{c}>90^{\circ}$, the orbital inclination of HD 73344 d is $I_{d}=121^{+17}_{-20}$ deg at $1\sigma$ confidence. On the other hand, for solutions $I_{c} < 90^{\circ}$, we get $I_{d}=59^{+19}_{-17}$ deg at $1\sigma$ confidence. Our results also show that HD 73344 is a Jupiter analog with a mass of $m_{d}=2.55^{+0.56}_{-0.46}\ \mathrm{M_{J}}$ and semi-major axis of $6.70^{+0.25}_{-0.26}$ AU.

\section{3D Architecture of HD73344 system}
\subsection{The mutual inclination between transiting planet and the outer giant planet}

\begin{figure}
    \centering
    \includegraphics[width=\linewidth]{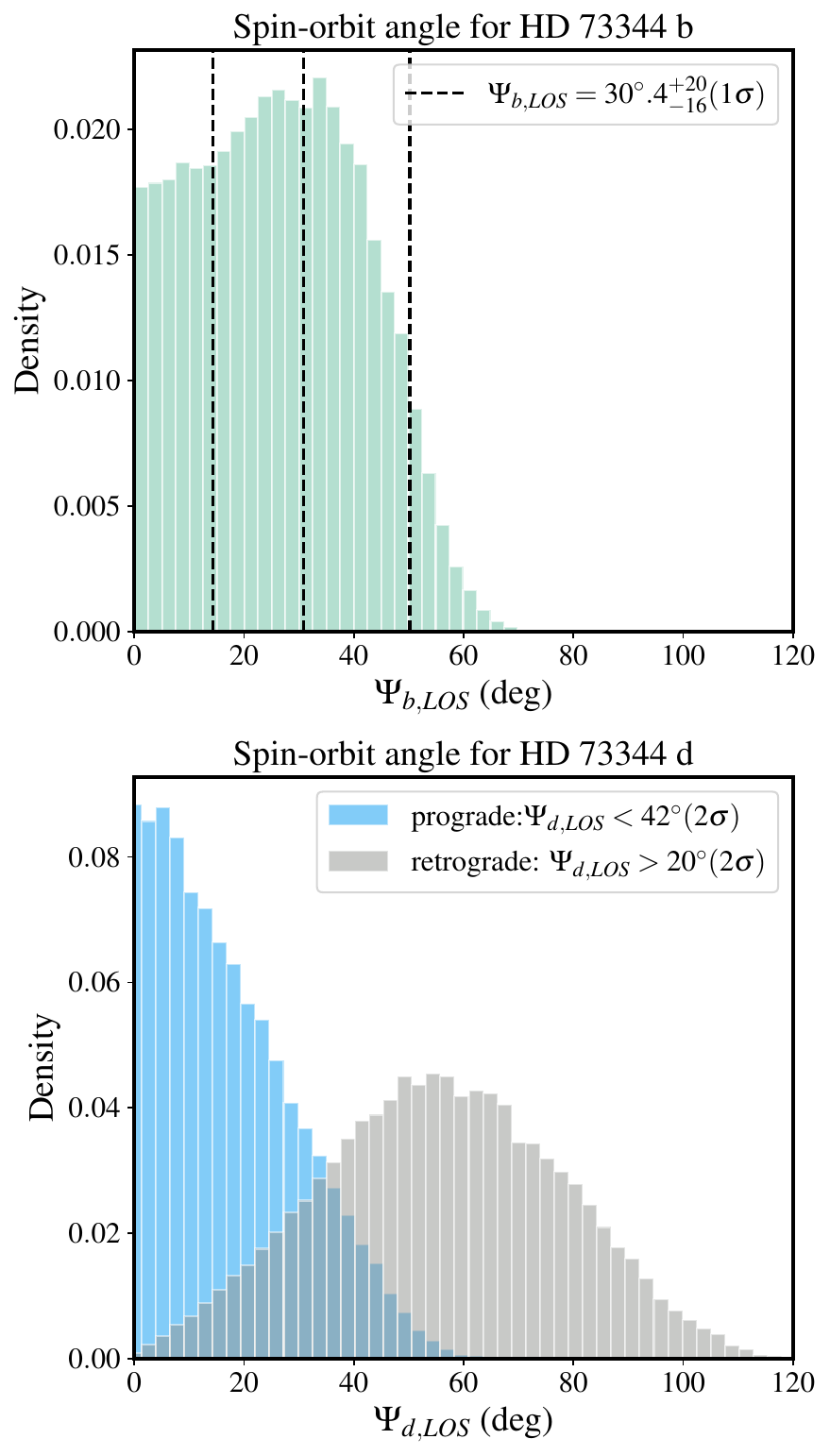}
    \caption{Top: the distribution of stellar obliquity projected in the line-of-sight direction for HD 73344 b. The black dashed lines correspond to max likelihood value with $1\sigma$ credible intervals. Bottom: the distribution of stellar obliquity projected in the line-of-sight direction for HD 73344 d for prograde (blue) and retrograde (gray) scenarios.}  
    \label{fig:figureobl}
\end{figure}

The transit observations constrain the orbital inclination of the inner-most planet b to be  nearly $90^{\circ}$ \citep{Sulis2024}. In contrast, our analysis of RVs and astrometric acceleration show that the orbital inclination of the outer giant planet $I_d$ is away from $90^{\circ}$. The difference suggests that the HD 73344 b and d are misaligned in line-of-sight direction. 
To better constrain their misalignment, we calculated the mutual inclination between HD 73344 b and d as \citep{Fabrycky2009}:  
\begin{equation}
    \cos{\Delta I_{bd}} = \cos{I_b}\cos{I_d} + \sin{I_b}\sin{I_d}\cos{(\Omega_b - \Omega_d)}
    \label{eq:Imut}
\end{equation} 
where $I_i$ and $\Omega_i$ is the inclination and longitude of ascending node for planet $i$. In our case,  we randomly sampled $I_{d}$ and $\Omega_{d}$ from the chains of our MCMC fitting and $I_{b}$ from a normal distribution $\mathcal{N}(88^{\circ}.48,0^{\circ}.05)$ from the transiting observations \citep{Sulis2024}. The only unknown parameter is $\Omega_{b}$ because we do not know the orientation of the inner orbit from transits and RVs. Thus, we randomly sample  $\Omega_{b}$ from a uniform distribution between 0 and $2\pi$. 

Figure~\ref{fig:figure6} presents the resulting distribution of the mutual inclination $\Delta I_{bd}$, which appears bimodal and symmetric around $90^{\circ}$.
As we have discussed in section~\ref{sec:3dfit}, planets orbiting in the same plane may orbit in opposite the direction, causing the prograde-retrograde degeneracy. Therefore, if HD 73344 b and d orbit the star in the same direction, their mutual inclination will be less than $90^{\circ}$. On the contrast, they will have a mutual inclination larger than $90^{\circ}$ if orbiting in opposite directions. Note that the double peaks in the mutual inclination distribution do not arise solely from the bimodal distribution of outer giant planet's orbital inclination $I_d$. They also arise because we do not know the moving direction of the transiting planet. Even if the distriubtion of $I_d$ values had only a single peak, we would still observe double peaks in the mutual inclination distribution.  We also noticed a relatively high probability for mutual inclinations around $90^\circ$, which corresponds to orthogonal orbits between HD 73344 b and d. This may not reflect reality since such orbits are rare; it likely results from the uniform distribution of $\Omega_b$ that we used. The mutual inclination is $46^{\circ}<\Delta I_{bc}<134^{\circ}$ at $1\sigma$ level and $20^{\circ}<\Delta I_{bc}<160^{\circ}$ at $2\sigma$ level. Therefore, we find a $>20^{\circ}$ misalignment between HD 73344 b and c ($2\sigma$). Our results strongly disfavor a coplanar architecture for HD 73344 system.

\subsection{The spin-orbit angle}\label{sec:obl}

In this paper, we define the spin-orbit angle as the angle between the stellar spin axis and the orbital axis of a planet. The top panel of Figure~\ref{fig:figureobl} presents the posterior distribution of spin-orbit angle for HD 73344 b projected in the line-of-sight direction. We combined the stellar inclination measurements from section~\ref{sec:sr_si} and the orbital inclination of HD 73344 b from transit observations \citep{Sulis2024}. Specifically, we computed the line-of-sight spin-orbit angle by taking the difference between the two inclinations. The distribution peaks at $30^{\circ}$ and suggests that HD 73344 b is misaligned relative to the host star ($1\sigma$ significance). Rossiter-McLaughlin (RM) observations of HD 73344 will provide the  spin-orbit angle projected in the sky plane and provide better constrained the spin-orbital misalignment of the transiting planet HD 73344 b. With a planet radius of $\sim2.88 R_{\oplus}$  and a stellar rotation velocity of $v\sin i_* \sim 5.67\ km\ s^{-1} $, the Rossiter-McLaughlin (RM) effect amplitude for HD 73344 b is estimated to be $2.5 m\ s^{-1}$. Observing the RM effect would require extreme-precision RV measurements. Additionally, stellar activity presents a significant challenge, as \citet{Sulis2024} reported stellar activity-induced RV jitter of approximately $\sim 4.7 m\ s^{-1}$.

The bottom panel of Figure~\ref{fig:figureobl} presents the posterior distributions of spin-orbit angle for HD 73344 d in the line-of-sight direction. Due to the bimodal feature of $I_{d}$, two distinct distributions emerge for the line-of-sight spin-orbit angle of HD 73344 d, corresponding to prograde and retrograde scenarios, respectively. In the prograde scenario, HD 73344 d exhibits spin-orbit angle consistent with an aligned orbit, with values  $<42^{\circ}$ at the $2\sigma$ confidence level. By contrast, in the retrograde scenario, the spin-orbit angle projected in line-of-sight direction is $>20^\circ$ (2$\sigma$).

\section{Orbital Dynamics of HD 73344 system}

\subsection{Strong coupling between two inner planets}\label{sec:coupling}
In this section, we investigate how the mutual inclination between two inner planets  HD 73344 b $\&$ c evolve under the influence from the outer giant planet. We first applied the the analytical model from \cite{DongandPu2017} to HD 73344 system. Specifically, we consider the orbital angular momenta of three planets as $\vec{L_{b}}=L_{b}\hat{l_{b}}$, $\vec{L_{c}}=L_{c}\hat{l_{c}}$ and $\vec{L_{d}}=L_{d}\hat{l_{d}}$, where $\hat{l_{b}}$, $\hat{l_{c}}$,and $\hat{l_{d}}$ are unit vectors along the orbital  axes of the planets. The directions of $\hat{l_{b}}$, $\hat{l_{c}}$ and $\hat{l_{d}}$ indicate the orientation of the planets' orbits. Because HD 73344 d has much larger mass and orbital distance than the other two planets ($m_{b}, m_{c}\ll m_{d}$, $a_{b}, a_{c}\ll a_{d}$), its orbits and angular momentum would be hardly affected by $\hat{L_{b}}$ and $\hat{L_{c}}$ and is approximately invariant. Therefore, we only consider the orbital evolution of HD 73344 b and c, which can be described as the precession of $\hat{l_{b}}$ and $\hat{l_{c}}$ around each other, and their precession around $\hat{l_{d}}$. We used $\nu_{bc}$ and $\nu_{bd}$ to represent the precession rate of $\hat{l_{b}}$ around $\hat{l_{c}}$ (HD 73344  b around HD 73344 c) and that of $\hat{l_{b}}$ around $\hat{l_{d}}$ (HD 73344  b around HD 73344 d). Similarly, $\nu_{cb}$ and $\nu_{cd}$ are the precession rates of $\hat{l_{c}}$ around $\hat{l_{b}}$ (HD 73344  c around HD 73344 b) and around $\hat{l_{d}}$ (HD 73344 c around HD 73344 d). \cite{DongandPu2017} use the parameter $\epsilon=(\nu_{bd}-\nu_{cd})/ (\nu_{bc}+\nu_{cb})$ to estimate the relative coupling strength between the inner planets compared to the `separate' force from the outer planet. If $\epsilon \gg 1 $, the two inner planets precess around the outer giant planet at different rates, which leads to de-coupling and large mutual inclination between inner planets. On the contrary, if $\epsilon \ll 1 $, the fast precession of the two inner planets around each other keeps them strongly coupled together.

For the HD 73344 system, we obtained $\nu_{bc}\approx  1.75 \ \rm{rad}\ \rm{kyr}^{-1}$, $\nu_{bd}\approx 0.003 \ \rm{rad}\ \rm{kyr}^{-1}$, $\nu_{cb}\approx 0.02 \ \rm{rad}\ \rm{kyr}^{-1}$, and $\nu_{cd}\approx 0.01 \ \rm{rad}\ \rm{kyr}^{-1}$. We derived  $\epsilon=0.0045 \ll1$, indicating the HD 73344 b and c are strongly coupled with each other. Therefore, the torque from the outer giant planet is not expected to excite a high mutual inclination between the inner planets.

Furthermore, we observe that the precession rate of HD 73344 b around c is significantly faster than the other three values ($\nu_{bc} \gg \nu_{bd}, \nu_{cb}, \nu_{cd}$). This is due to HD 73344 c having a mass similar to Saturn and being much closer to the star than HD 73344 d, causing its gravitational influence on HD 73344 b to be dominant over that from HD 73344 d.

\begin{figure*}
    \centering
    \includegraphics[width=\linewidth]{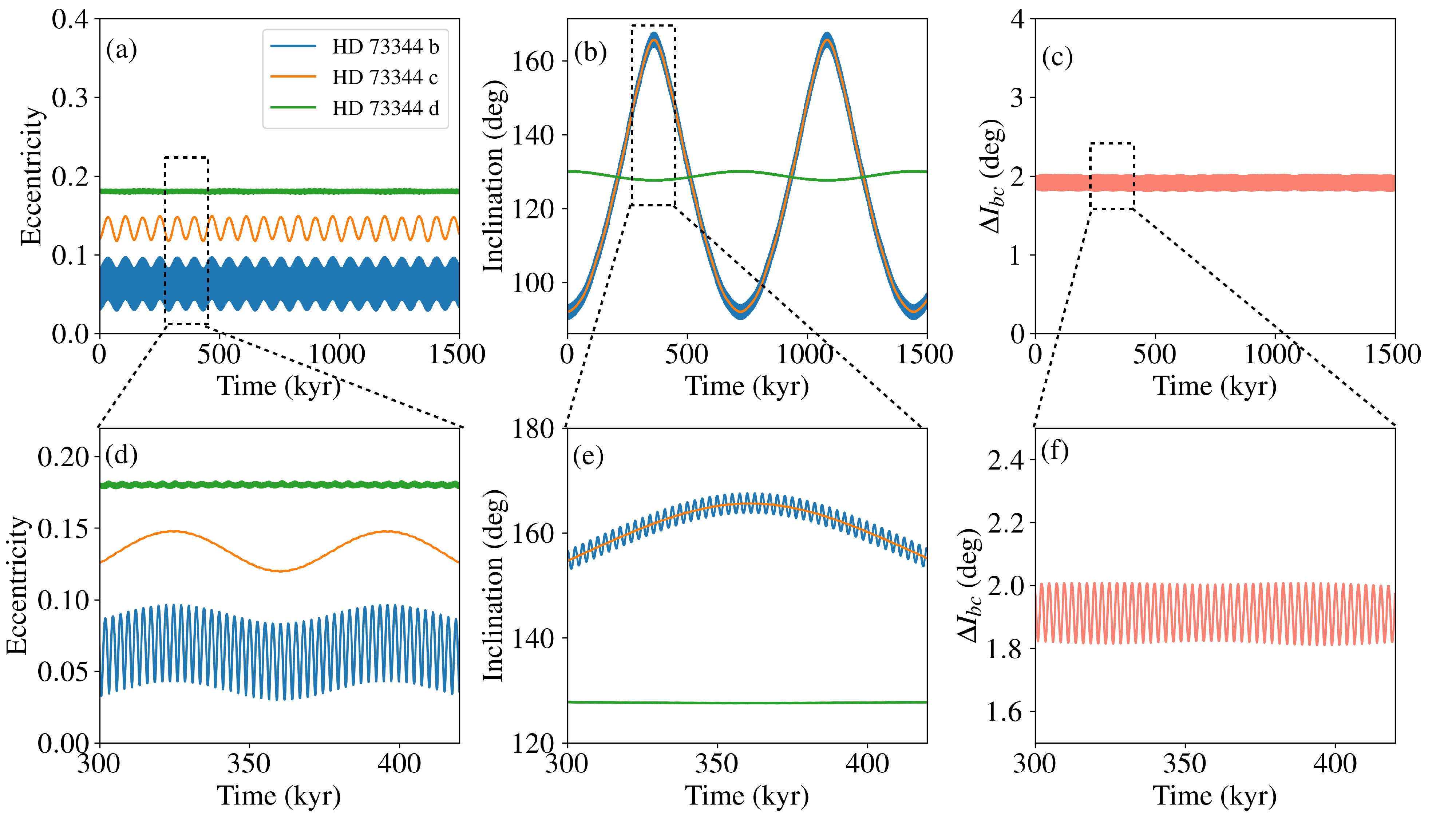}
    \caption{N-body simulations of dynamical evolution of HD 73344 system. (a): The eccentricities of three planets as the function of time. (b): The orbital inclination of the three planets as the function of time. (c): The mutual inclination between two inner planets as the function of time. (d)-(f): the zoom-in version of panel (a)-(c). }  
    \label{fig:figure_sim}
\end{figure*}

Figure~\ref{fig:figure_sim} presents the N-body simulation using package \texttt{REBOUND} \citep{Rein2012} with initial conditions of planets from the max likelihood values from Table~\ref{tab:rvparams} and Table~\ref{tab:3d}. We set a low initial mutual inclination of $2^{\circ}$ between HD 73344 b and c, assuming they were nearly coplanar at formation, yet with a mutual inclination still large enough for only the innermost planet to transit. We set the outer giant planet to be misaligned relative to HD 73344 b by $40^{\circ}$. We also set the initial longitude of ascending node ($\Omega$) for all planets to be zero. In addition, we incorporated the general relativity (GR) effects from the host star using the \texttt{REBOUNDx} \citep{reboundx2020} in the simulation. The N-body results were consistent with predictions from the analytical model: HD 73344 b and c precess about each other at a much faster rate than their precession around the outer giant planet, d. During the precession, the inclination of HD 73344 b and c oscillate as a function of time while their mutual inclination remains low, oscillating between 1.8 and 2.0 degrees. 

Furthermore, we conducted two additional N-body simulations, keeping all other parameters same as above simulation but varying the initial mutual inclination between HD 73344 b and c. When the two inner planets have a moderate mutual inclination ($I_b=90^{\circ}, I_c = 110^{\circ}, \Delta I_{bc}=20^{\circ}$),  the system remains in a stable configuration. In this scenario, the inner planets precess together around the outer giant planet (see Figure~\ref{fig:figure_sim2}). The mutual inclination $\Delta I_{bc}$ stays approximately $20^{\circ}$ throughout the simulation. In contrast, if we set the initial $\Delta I_{bc}$ as $70^{\circ}$ ($I_b=90^{\circ}, I_c = 160^{\circ})$, the system becomes highly unstable, with the eccentricity of the innermost planet rapidly increasing to 1 (see Figure~\ref{fig:figure_sim3}). This is consistent with the dynamical analysis in \cite{Sulis2024}, which examines the stability of the system without accounting for the outer giant planet. Their analysis indicate that only large mutual inclinations ($>60^{\circ}$) between the two inner planets are incompatible with long-term stability. 
Therefore, we cannot rule out the possibility of a moderate mutual inclination between the inner planets. 

Meanwhile, with an orbital distance of $\ttilde 0.33$ AU, the angle range for a transiting configuration of HD 73344 c is only $\ttilde 1.7^{\circ}$. Therefore, the fact that  HD 73344 c is not transiting provides a lower limit of $1.7^{\circ}$ for the mutual inclination between HD 73344 b and c. 

%Our dynamical analysis indicates a strong coupling between HD 73344 b and c. However, HD 73344 c has not been observed transiting in \textit{K2} and \textit{TESS} data \citep{Sulis2024}. With an orbital distance of $\ttilde 0.33$ AU, the angle range for a transiting configuration of HD 73344 c is only $\ttilde 1.7^{\circ}$. Therefore, the fact that  HD 73344 c is not transiting provides a lower limit of $1.7^{\circ}$ for the mutual inclination between HD 73344 b and c.

%It is plausible that HD 73344 c has a low mutual inclination relative to HD 73344 d, but remains outside the transiting configuration.

\subsection{Spin-orbit evolution of inner planets}

We applied the results of the `three-vector problem' \citep{BL06,BL09,BF14I,BF14}, a model for the secular evolution of three coupled angular motions, to explore the evolution of the spin-orbit angle for inner planets. In section~\ref{sec:coupling}, we showed that the two inner planets strongly coupled with each other. Here, we consider the total angular momentum of the inner planetary system $\vec{L}_{in}$ instead of the two individual planets $\vec{L}_{in}=\sum_{j=b,c}L_{j}\hat{l}_{j}=L_{in}\hat{l}_{in}$, where $\hat{l}_{j}$ is the unit vector normal to the orbit of planet $j$. The other two vectors are the angular momentum of the star $\vec{L_{\star}}=L_{\star}\hat{l}_{\star}$, and that of the outer giant planet $\vec{L_{d}}=L_{d}\hat{l}_{d}$, where $\hat{l}_{\star} $ is the unit vector along stellar spin axis and $\hat{l}_{d}$ is the unit vector along the orbital axis of giant planet. Following the convention in \cite{BF14}, we denote $\nu_{1}$ as the precession rate of $\hat{l}_{\star}$ around $\hat{l}_{in}$, $\nu_{2}$ as the precession rate of $\hat{l}_{in}$ around $\hat{l}_{\star}$, $\nu_{3}$ as the precession rate of $\hat{l}_{in}$ around $\hat{l}_{d}$, and $ \nu_{4}$ as the precession rate of $\hat{l}_{d}$ around $\hat{l}_{in}$, respectively. 

For HD 73344 , we estimate $\nu_{1}\approx \ 1.8\times10^{-5} \ \rm{rad}\ \rm{kyr}^{-1}$, $\nu_{2}\approx \ 7.0\times10^{-6} \ \rm{rad}\ \rm{kyr}^{-1}$, $\nu_{3}\approx \ 8.4\times10^{-3} \ \rm{rad}\ \rm{kyr}^{-1}$, $\nu_{4}\approx \ 2.4\times10^{-4}\ \rm{rad}\ \rm{kyr}^{-1}$. We find the HD 73344 system is consistent with the ``Pure Orbital Regime'' in \citet{BF14}, with $\nu_{1},\nu_{2},\nu_{4}\ll \nu_{3}$. In this regime, the precession rate associated with the inner planetary system coupling to the stellar spin ($\nu_{1}$ and $\nu_{2}$) are much smaller than those associated with the inner planetary system coupling to the outer planet ($\nu_{3}$ and $\nu_{4}$). This suggests that the stellar spin would neither significantly influence the orbits of planets nor be affected by the motion of planets. Meanwhile, $\nu_{4}\ \ll\ \nu_{3} $ suggests that the orbital plane of HD 73344 d is almost fixed ($\nu_{4}\ \ll\ \nu_{3} $). Therefore, the dominant evolution in HD 73344 is the precession of inner planets around the orbital axis of the outer giant planet at a roughly constant angle. The precession period is $P=2\pi/\nu_{3}\approx 740$ kyr, which is confirmed in our N-body simulations (see Figure \ref{fig:figure_sim}). In Section~\ref{sec:obl}, our analysis reveals that HD 73344 b is likely misaligned with the stellar spin axis in light of sight direction. This misalignment could be explained by the precession induced by the outer giant planet \citep{BF14,zhang2021}. Consequently, we anticipate that Rossiter-McLaughlin (RM) observations of the transiting planet will detect a misalignment between the planet's orbit and the stellar spin axis, and the amount of this misalignment will provide additional constraints on the inclination of outer giant planet d.

\section{Discussion}

\subsection{Possible Mechanisms to Cause the inner-outer misalignment }

HD 73344 joins a small but growing list of exoplanet systems hosting close-in small planets that are found to be misaligned relative to a distant giant planet using RV and astrometry, e.g. HAT-P-11 \citep{Winn2010,Yee2018}, $\pi$ Men \citep{Jones2002,Kunovac2020,Xuan2020,DeRosa2020}. Additionally, Kepler-56 \citep{Huber2013, Otor2016} and Kepler-129 \citep{zhang2021} are found to be misaligned relative to the spin axes of their host stars, likely due to the influence of the outer giant planets in the systems. These systems stand as a stark contrast to the flatness of our own solar system. It is natural to ask: what causes the mutual inclination between the inner and outer planets in these systems?

One possible scenario is that all planets form in the protoplanetary disc and are aligned with each other initially. At some point, two or more giant planets undergo dynamical encounters and only one giant planet survives, ending up with a high inclination orbit relative to the initial disk plane \citep{Rasio1996,Chatterjee2008,Beaug2012, Petrovich2014}. The giant planets in HAT-P-11 and $\pi$ Men are consistent with this mechanism, as they have high eccentricities ($e\sim 0.6$) and high inclinations. But HD 73344 d, along with the giant planets in Kepler-56 and Kepler 129 systems, all have a low eccentricity ($e \lesssim 0.2$). It is also still an open question whether the giant planet could attain a mutual inclination up to $40^{\circ}$ through planet-planet scattering. 

Another possibility is that the planets formed in a warped protoplanetary disc with misaligned inner and outer components \citep{zanazzi2018}. In such a case, the inner and outer components of protoplanetary disk could be misaligned due to perturbations from a misaligned companion \citep{nealon2019}. In HD 73344, the obvious candidate is the giant planet HD 73344 d, but this would require the giant planet to start out misaligned relative to the disk first. Alternatively, the misalignment between inner and outer disks could be caused by  perturbations
from a fly-by star. However, \cite{Nealon2020} found that a stellar flyby may be too short-lived to cause a significant effect. Although the maligning mechanism is till unclear, recent high-resolution observations of transition disks using Atacama Large Millimeter/submillimeter Array (ALMA) have reveals the existence of  misalignment between the inner and outer disks \citep{Bohn2022} .  the planet-planet scattering may produce relatively high eccentricities for the remaining giant planets whereas a warped protoplanetary disc does not, more discoveries of misaligned planetary systems in the future can help to study the eccentricity distribution of the outer giant planets and statistically and distinguish between these two formation mechanisms.

%($m_{d}=2.55^{+0.56}_{-0.46}\ \mathrm{M_{J}}$, $a_{d}=6.70^{+0.25}_{-0.26}$ AU, $e_{d}=0.18^{+0.14}_{-0.12}$) based on 27-year radial velocity observations from ELODIE, Lick/Hamilton, SOPHIE, APF and Keck/HIRES. HD 73344 also hosts a compact inner planetary system, including a transiting sub-Neptune HD 73344 b ($P_{b}=15.61\ \mathrm{days}$, $r_{b}=2.88^{+0.08}_{-0.07}\ \mathrm{R_{\oplus}}$) and a non-transiting Saturn-mass planet ($P_{c}=65.936\ \mathrm{days}$, $m_{c}\sin{i_c}=0.367^{+0.022}_{-0.021}\ \mathrm{M_{J}}$).  By analyzing  \textit{TESS} light curves, we identified a stellar rotation period of $9.03\pm{1.3}$ days.

\section{Conclusion}

We have presented the discovery of a long-period giant planet outside two small close-in planets in the HD 73344 system. We have measured the mutual inclination between the innermost planet and the outer giant planet, and studied the orbital dynamics of the system. Our main conclusions are as follows:

\begin{itemize}
    \item We confirm that the nearby transiting planet host star HD 73344 also hosts a non-transiting Saturn-mass planet HD 73344 c ($P_{c}=65.936\ \mathrm{days}$, $m_{c}\sin{i_c}=0.367^{+0.022}_{-0.021}\ \mathrm{M_{J}}$, $e_c=0.124^{+0.052}_{-0.053}$) consistent with the discovery of \cite{Sulis2024}.
    
    \item  We identified a stellar
rotation period of $9.03\pm{1.3}$ days from \textit{TESS} light curves. Combining with spectroscopically determined stellar rotation speed $v\sin{i_*}$, we constrained the stellar inclination to be $i_*=63^{\circ}.6^{+17.4}_{-16.5}$. The result suggests that the transiting innermost planet HD 73344 b may be misaligned relative to the stellar spin axis ($1.5\sigma$ significance).  Rossiter-McLaughlin observations are ideally poised to strengthen (or refute) this result.

\item We discovered a Jupiter-like planet HD 73344 d ($m_{d}=2.55^{+0.56}_{-0.46}\ \mathrm{M_{J}}$, $a_{d}=6.70^{+0.25}_{-0.26}$ AU, $e_{d}=0.18^{+0.14}_{-0.12}$) outside the compact inner planetary system. We constrained the orbital inclination of HD 73344 d ($I_{d}=121^{+17}_{-20}$ deg for $I_{d}>90^{\circ}$ and $I_{d}=59^{+19}_{-17}$ deg for $I_{d}<90^{\circ}$ ) by combing RV and Hipparcos-Gaia astrometric accelerations. The mutual inclination between HD 73344 b and d is $46^{\circ} <\Delta I_{bd}< 134^{\circ}\ (1\sigma)$ and $20^{\circ} <\Delta I_{bd}< 160^{\circ}\ (2\sigma)$, strongly disfavoring coplanar architectures. 

    \item The analytical model and N-body simulation both show that the two inner planets strongly couple with each other while they precess around the orbit normal of the inclined, outer giant together. During the precession, the orbital inclination of inner planets oscillates with time, allowing for observations in which these planets are misaligned with respect to the spin axis of the host star.
    
\end{itemize}

%one more paragraph here about significance and/or future outlook
With the upcoming release of Gaia DR4 time-series astrometry, we will be able to measure the mutual inclination of outer giant planets relative to the inner planetary systems around more stars like HD 73344. This will enable us to analyze the flatness of exoplanetary systems on a statistical level.

\acknowledgments

The authors wish to recognize and acknowledge the very significant cultural role and reverence that the summit of Maunakea has always had within the indigenous Hawaiian community. We are most fortunate to have the opportunity to conduct observations from this mountain.

L.M.W. acknowledges support from the NASA Exoplanet Research Program (grant no. 80NSSC23K0269). J.Z. would like to thank Jason Wang for  helpful advice regarding the use of package \texttt{PyKlip}. J.Z. and D.H. acknowledge support from the Alfred P. Sloan Foundation, the National Aeronautics and Space Administration (80NSSC22K0303).  D.H. also acknowledges support from the Alfred P. Sloan Foundation, and the Australian Research Council (FT200100871). N.S. acknowledges support by the National Science Foundation Graduate Research Fellowship Program under Grant Numbers 1842402 \& 2236415.

%This publication makes use of The Data & Analysis Center for Exoplanets (DACE), which is a facility based at the University of Geneva (CH) dedicated to extrasolar planets data visualisation, exchange and analysis. DACE is a platform of the Swiss National Centre of Competence in Research (NCCR) PlanetS, federating the Swiss expertise in Exoplanet research. The DACE platform is available at https://dace.unige.ch.

\vspace{5mm}
\facilities{Keck(HIRES), Lick(Halminton), APF, SOPHIE, ELODIE, TESS, Gaia, Hipparcos}

%% Similar to \facility{}, there is the optional \software command to allow 
%% authors a place to specify which programs were used during the creation of 
%% the manuscript. Authors should list each code and include either a
%% citation or url to the code inside ()s when available.

\software{ \textit{RadVel} \citep{Fulton2018} \textit{TESS-SIP}\citep{Hedges2020} \textit{PyKlip}\citep{Wang2015} \textit{REBOUND}\citep{Rein2012} \textit{REBOUNDx}\citep{reboundx2020}}

%% Appendix material should be preceded with a single \appendix command.
%% There should be a \section command for each appendix. Mark appendix
%% subsections with the same markup you use in the main body of the paper.

%% Each Appendix (indicated with \section) will be lettered A, B, C, etc.
%% The equation counter will reset when it encounters the \appendix
%% command and will number appendix equations (A1), (A2), etc. The
%% Figure and Table counter will not reset.

\newpage
\counterwithin{figure}{section}
\appendix
\section{Supplementary figures}
\begin{figure*}
    \centering
    \includegraphics[width=\linewidth]{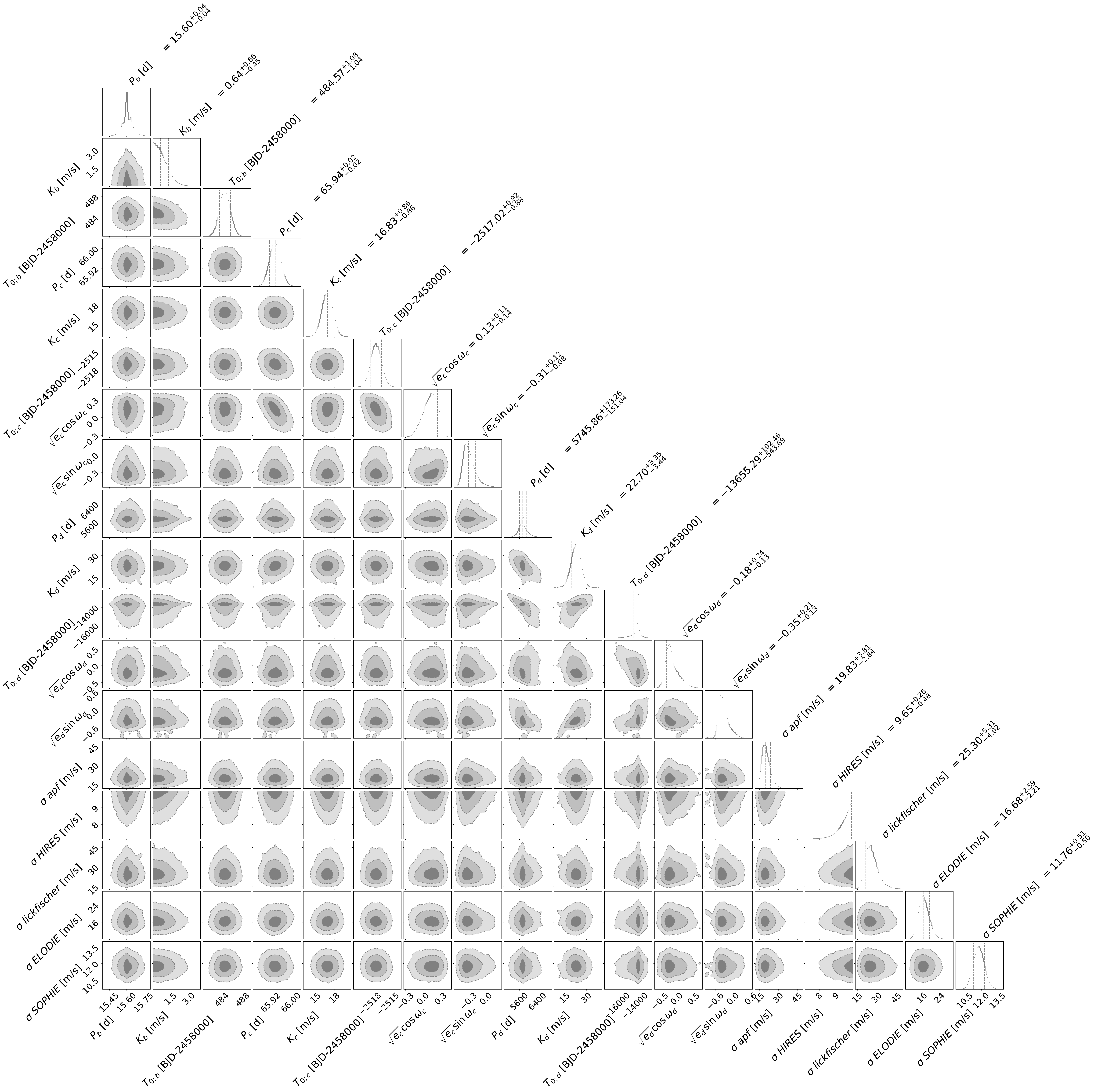}
    \caption{ Joint posterior distributions for 16 parameters used in the RV-only fitting of a 3-planet Keplerian orbital model for HD 7334 system.  }  
    \label{fig:figureA1}
\end{figure*}

\newpage

\begin{figure*}
    \centering    \includegraphics[width=\linewidth]{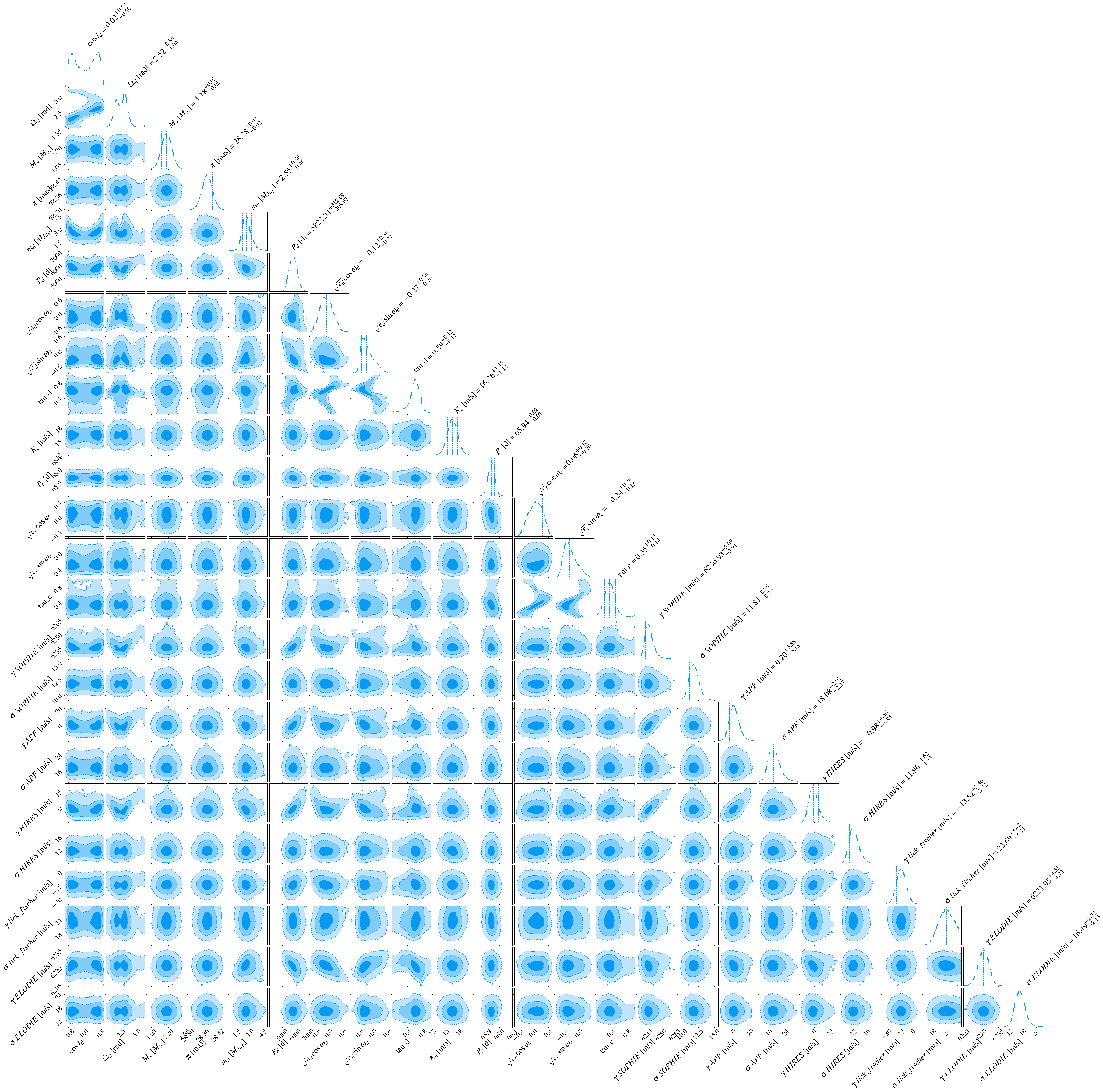}
    \caption{ Joint posterior distributions for 22 parameters used in the MCMC fitting for the 3D orbits of the outer planet HD 73344 d. The values and histogram distributions of all parameters are shown, along with 1 $\sigma$ uncertainties. }  
    \label{fig:figureA2}
\end{figure*}

\newpage

\begin{figure*}
    \centering
    \includegraphics[width=\linewidth]{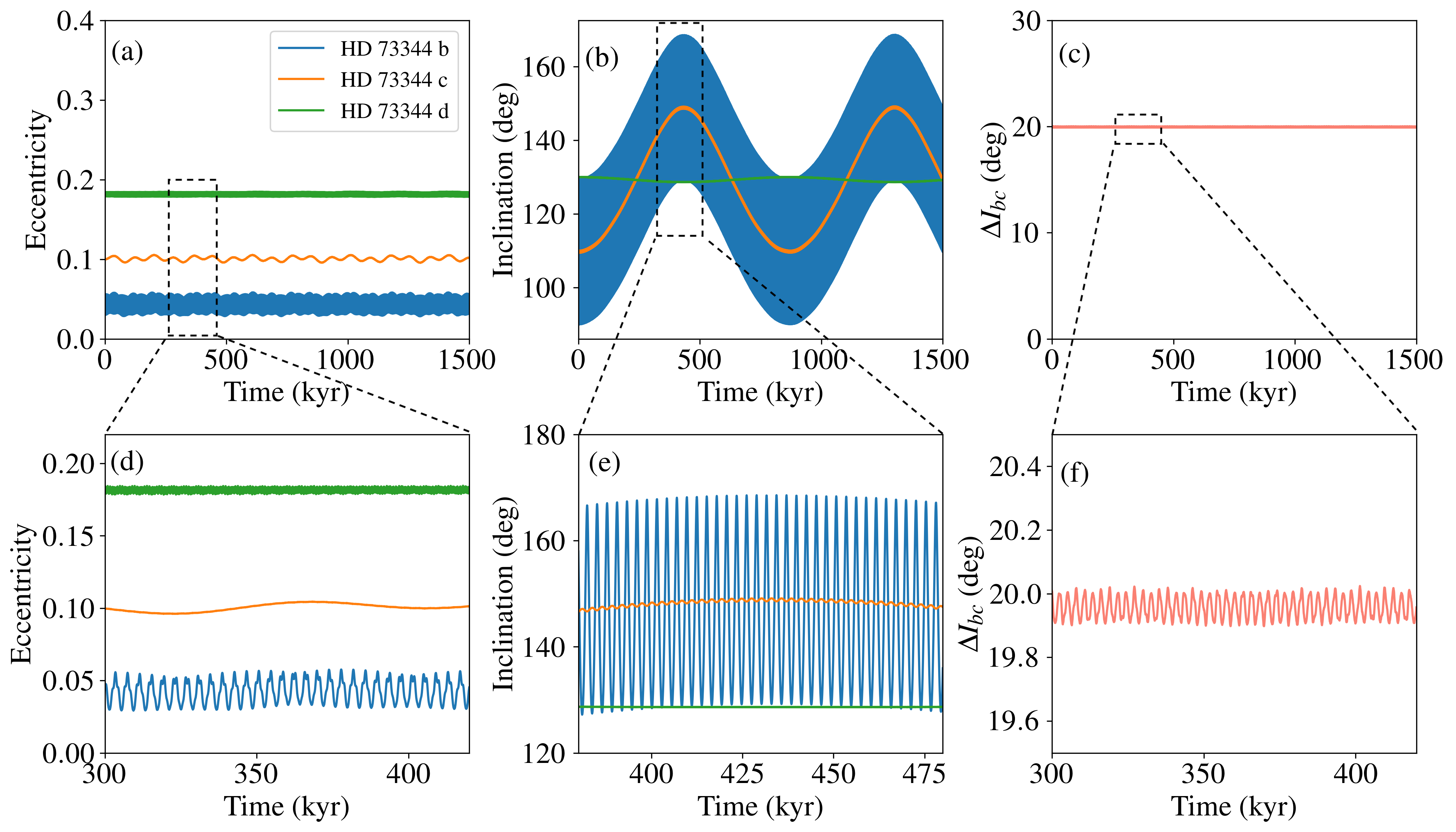}
    \caption{Same as Figure~\ref{fig:figure_sim}, but with an initial mutual inclination of $20^{\circ}$ between HD 73344 b$\&$c. }  
    \label{fig:figure_sim2}
\end{figure*}

\begin{figure*}
    \centering
    \includegraphics[width=\linewidth]{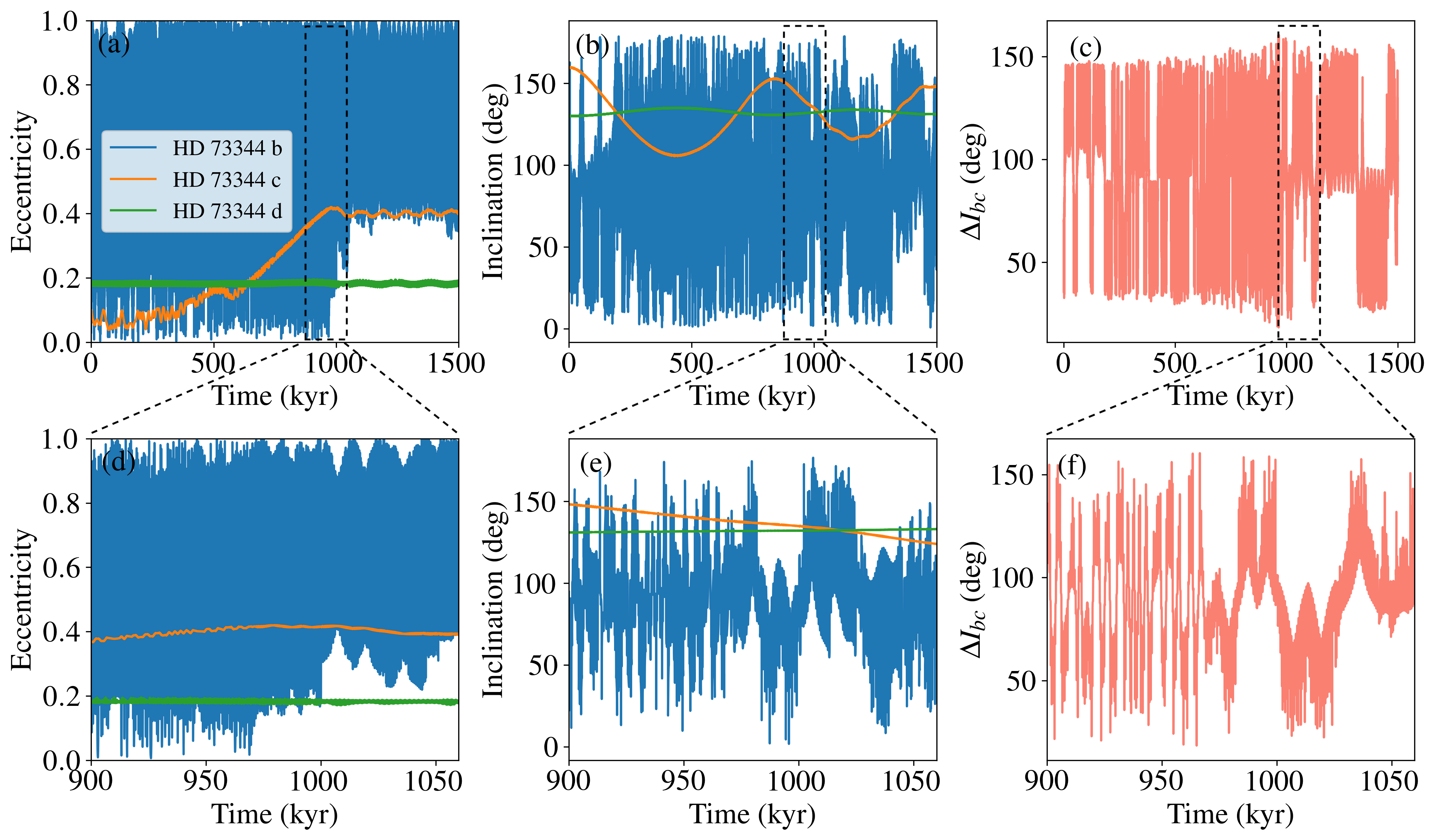}
    \caption{Same as Figure~\ref{fig:figure_sim}, but with an initial mutual inclination of $70^{\circ}$ between HD 73344 b$\&$c. }  
    \label{fig:figure_sim3}
\end{figure*}

%% For this sample we use BibTeX plus aasjournals.bst to generate the
%% the bibliography. The sample63.bib file was populated from ADS. To
%% get the citations to show in the compiled file do the following:
%%

\bibliography{sample63}{}
\bibliographystyle{aasjournal}

%% This command is needed to show the entire author+affiliation list when
%% the collaboration and author truncation commands are used.  It has to
%% go at the end of the manuscript.
%\allauthors

%% Include this line if you are using the \added, \replaced, \deleted
%% commands to see a summary list of all changes at the end of the article.
%\listofchanges
\end{CJK*}
\end{document}